\documentclass[preprint,showpacs,preprintnumbers,amsmath,amssymb,natbib]{revtex4}

\usepackage{graphicx}
\usepackage{epsfig}		
\usepackage{dcolumn}
\usepackage{bm}
\def\x{\mbox{s}}
\DeclareMathOperator{\sech}{sech}
\DeclareMathOperator{\arctanh}{arctanh}
\DeclareMathOperator{\arccosh}{arccosh}

\begin{document}

\title{Cyclically deformed defects and topological mass constraints}

\author{A. E. Bernardini}
\email{alexeb@ufscar.br, alexeb@ifi.unicamp.br}
\affiliation{Departamento de F\'{\i}sica, Universidade Federal de S\~ao Carlos, PO Box 676, 13565-905, S\~ao Carlos, SP, Brasil}
\author{Rold\~ao da Rocha}
\email{roldao.rocha@ufabc.edu.br}
\affiliation{Centro de Matem\'atica, Computa\c c\~ao e Cogni\c c\~ao,
Universidade Federal do ABC 09210-170, Santo Andr\'e, SP, Brazil}

\date{\today}

\begin{abstract}
A systematic procedure for obtaining defect structures through cyclic deformation chains is introduced and explored in detail.
The procedure outlines a set of rules for analytically constructing constraint equations that involve the finite localized energy of cyclically generated defects.
The idea of obtaining cyclically deformed defects concerns the possibility of regenerating a primitive (departing) defect structure through successive, unidirectional, and eventually irreversible, deformation processes.
Our technique is applied on kink-like and lump-like solutions in models described by a single real scalar field, such that extensions to quantum mechanics follow the usual theory of deformed defects.
The preliminary results show that the cyclic device  supports simultaneously kink-like and lump-like defects into $3$- and $4$-cyclic deformation chains with topological mass values closed by trigonometric and hyperbolic deformations.
In a straightforward generalization, results concerning with the analytical calculation of $N$-cyclic deformations are obtained and lessons regarding extensions from more elaborated primitive defects are depicted.
\end{abstract}

\date{\today}
\maketitle

\section{Introduction}

Solutions of a restricted class of nonlinear equations yield an ample variety of topological (kink-like) and non-topological (lump-like) defects of current interest, in particular to mathematical and physical applications as well \cite{BookA}.
More precisely, these defect structures are finite-energy solutions of a nonlinear partial differential equation,
 which tend to be stabilized by a conserved charge associated with an underlying field theory \cite{BookB}.
To get a more clarifying and effective definition, the equation which leads to the corresponding classical solution needs two main terms: a sharpening nonlinear term and a dispersive term.
Essentially, the energy density of the localized solution captures the nonlinear nature of the corresponding Lagrangian, making its dynamics an insightful topic.
In this sense, topological and non-topological structures are not only of mathematical interest.

Defect structures are indeed investigated in cosmological scenarios involving seeds for structure formation \cite{0001,0002,0003,0003B}, Q-balls \cite{0004,0004B}, and tachyon branes \cite{0005,0006,0007}; in braneworld scenarios with a single extra-dimension of infinity extent \cite{Bas05,Bas05B,0008,0009}; in particle physics as magnetic monopoles \cite{0010}; and in some applications on soft condensed matter physics \cite{Books} involving, for instance, charge transport in diatomic chains \cite{0014,0015,0016}, or even baby {\em skyrmions} \cite{0010B}.
Specific studies involving kink-like defects are also used to be related to the symmetry restoration in inflationary scenarios \cite{0012,0011}, or to the fermion effective mass generation induced by some symmetry breaking mechanism \cite{0017,0018}.
Likewise, applications of current interest predominantly include the use of lump-like models to describe the behavior of bright solitons in optical fibers \cite{0017A,0018A}.
In particular, it represents the most prominent candidate for ultra-high-speed packet-switched optical networks in the current communication revolution.

Notwithstanding the pop science status of this field -- once its proponents seek to put it to use for transporting vast amounts of information -- the analysis regarding defect structure solutions may still require the introduction of challenging mathematical devices amalgamating analytical, geometrical, and sometimes complex numerical protocols.
Therefore, finding a systematic process for obtaining cyclically deformed defects, constrained by closure relations involving their topological masses, corresponds to the well succeeded purpose of this work.
Hereon the nomenclature of topological mass is arbitrarily extended to encompass the finite total energy of non-topological lump-like solutions.

Our work is concerned with a systematic procedure for defect generation defined through trigonometric and hyperbolic bijective deformation functions. They allow the construction of $N$-finite cyclic deformation chains involving topological and non-topological solutions.
The obtained defects are necessarily generated and regenerated from a primitive defect of an original theory (for instance, the $\lambda \phi^4$ theory).

In this context, several simplificative techniques have been suggested to study and solve non-linear equations through deformation procedures \cite{Bas05,Bas05B,Bas04}.
The main issues that have stimulated us to investigate cyclically deformed defects have indeed appeared through the BPS first order framework \cite{Bas01,Bas02,Bas03} for obtaining kink-like and lump-like structures.

Such an outstanding process of systematically obtaining cyclically deformed defects allows reedifying a preliminarily introduced defect structure - for instance, the usual kink defect from the $\lambda \phi^4$ theory - through successive, unidirectional, and eventually irreversible, deformation operations.
This possibility of regenerating a primitive defect is one of our principal concerns in this work.
Our second main contribution consists in systematically defining constraint relations, involving topological masses of the corresponding deformed defects which belong to the cyclically deformed chain.
Our technique also supports kink-like and lump-like solutions in models described by a single real scalar field, such that extensions to quantum mechanics follow the usual theory of deformed defects \cite{Bas01}.

The manuscript is therefore organized as follows.
As a preliminary illustration of the method, in section II we report about some known topological and lump-like solutions to identify the possibilities of finding some $3$-cyclic deformation chain, which speculatively engenders some constraints upon the topological masses related to the cyclically deformed defects.
Along this section we shall follow the BPS first order framework.
Our results for a second set of $3$-cyclic deformation allow one to identify a {\em plateau}-shaped form for the potential of a deformed kink-solution which, for instance, introduces the possibility of a kind of slow-rolling behavior for the scalar field deformed solution, $\varphi$, when its dynamics is turned on.
A systematic and generalized procedure that simultaneously supports kink-like and lump-like defects into $3$-cyclic deformation chains is introduced and scrutinized in section III.
The topological conserved quantities are analytically computed and the constraint relations, involving trigonometric and hyperbolic deformation functions, are straightforwardly obtained.
The results from section III are extended to a $4$-cyclic deformation device which is furthermore driven by hyperbolic and trigonometric deformation chains.
Our results shall ratify that the modified topological masses (or the total energy of localized solutions, in the case of lumps), computed for defect structures cyclically deformed from a triggering $\lambda \chi^4$ theory, are mutually constrained.
Finally, the hyperbolic and trigonometric deformation procedures are systematically generalized to $N$-finite cyclically deformed defects in section V.
We draw our conclusions in section VI by attempting to some important insights related to the study of defect structures which bring up the novel concept of $N$-cyclic deformations.

\section{Cyclically deformed defects - Non systematic procedure}

Assuming that the presence of first-order equations evidently simplifies the calculations,
along this manuscript we shall report about the BPS first order framework \cite{Bas01,Bas02,Bas03} for investigating topological (kink-like) and non-topological (lump-like) structures.

Let us firstly consider a set of three cyclically deformed defects, namely $\psi(\x)$, $\phi(\x)$, and $\varphi(\x)$, with $\x$ as the spatial coordinate, in a way that the corresponding $3$-cyclic deformation procedure can be prescribed by the following first-order coupled equations,
\begin{eqnarray}
\frac{d x}{d \varphi} \equiv  x_{\varphi} &=& \frac{d \varphi}{d \x} = y_{\phi} \varphi_{\phi}    = z_{\psi} \varphi_{\psi}, \nonumber\\
\frac{d y}{d \phi} \equiv  y_{\phi}    &=& \frac{d \phi}{d \x} = z_{\psi} \phi_{\psi}       = x_{\varphi} \phi_{\varphi}, \nonumber\\
\frac{d z}{d \psi} \equiv z_{\psi}    &=& \frac{d \psi}{d \x} = x_{\varphi} \psi_{\varphi} = y_{\phi} \psi_{\phi},
\label{topo02}
\end{eqnarray}
where the subindexes stand for the corresponding derivatives, with $\alpha_{\beta} = 1 / \beta_{\alpha}$ for
$\alpha,\, \beta =\, \phi,\, \varphi,\, \psi$, being identified with deformation bijective functions that generate novel families of deformed defects.
In this case, $ x_{\varphi}$, $y_{\phi}$, and $z_{\psi}$ are derivatives of the auxiliary superpotentials, and  the corresponding BPS form of the cyclically derived potentials shall be given by
\begin{equation}
U(\varphi) = \frac{1}{2} x_{\varphi}^2 ~\rightleftarrows ~
V(\phi) = \frac{1}{2} y_{\phi}^2 ~\rightleftarrows ~
W(\psi) = \frac{1}{2} z_{\psi}^2 ~\rightleftarrows ~
U(\varphi) = \frac{1}{2} x_{\varphi}^2,
\label{topo02B}
\end{equation}
such that the simplified framework of deformed defects \cite{Bas01} is straightforwardly recovered.
Such a sequence of $3$-cyclic deformations naturally obeys the chain rule constraint given by
\begin{equation}
\psi_{\phi} \phi_{\varphi} \varphi_{\psi}= 1.
\label{topo03}
\end{equation}

Due to simplificative reasons, the starting point of our discussion will be the {\em dimensionless} $\lambda \phi^4$ theory with a scalar potential given by
\begin{equation}
V(\phi) = \frac{1}{2} y_{\phi}^2  =  \frac{1}{2} (1 - \phi^2)^2, 
\label{topo04}
\end{equation}
from which, upon solving the equation of motion,
\begin{equation}
\frac{d^2}{d\x^2}\phi(\x) = \frac{d}{d\phi} V(\phi) = y_{\phi}\,y_{\phi\phi},
\label{topo05}
\end{equation}
one obtains the static solution described by a topological kink ($+$ sign) (or antikink ($-$ signal)) as
\begin{equation}
\phi(\x) = \pm \tanh{(\x)} ~~ \mbox{with} ~~ y_{\phi} = \sech{(\x)}^2,
\label{topo06}
\end{equation}
for which we shall follow the sign convention adopted in \cite{Bas02}, nevertheless reducing our analysis to $+$ sign solutions into the above equation.

As it is well-known, the potential $V(\phi)$ engenders one maximal point at $\phi^{Max} = 0$, and two critical points, $\phi^{0}_{\pm}$, which are also function {\em zeros}, i. e.
\begin{equation}
V(\phi^{0}_{\pm}) = 0 ~~\quad \mbox{and}\quad \left.\frac{d V}{d\phi}\right|_{\phi = \phi^{0}_{\pm}} = 0,
\end{equation}
that correspond to the asymptotic values of the kink solution,
\begin{equation}
\phi^{0}_{\pm} = \phi(\x\rightarrow\pm\infty) = \pm 1,
\end{equation}
namely its topological indexes.
The topological charge is correspondingly given by
\begin{equation}
Q^{\phi} = \left|\int^{+\infty}_{-\infty}{d\x \, y_{\phi}}\right| = \left|\phi^{0}_{+} - \phi^{0}_{-}\right| = 2,
\end{equation}
where any dimensional multiplicative factor was suppressed from the beginning, and the topological mass, which corresponds to the total energy of the localized solution, is given by
\begin{equation}
M^{\phi} = \left|\int^{+\infty}_{-\infty}{d\x \, y_{\phi}^2}\right| = 4/3,
\end{equation}
where the corresponding localized energy density has been identified with
\begin{equation}
\rho(\phi(\x)) = y_{\phi}^2 = \sech{(\x)}^4.
\end{equation}

At first glance, besides Eq.~(\ref{topo05}), our description of cyclically deformed defects could follow no additional systematic constraints on the choice of $\psi(\x)$, $\phi(\x)$, and $\varphi(\x)$.
Just as a preliminary illustration of the method, let us firstly obtain two sets of cyclically deformed defects from the  primitive kink solution above.

\subsection{First case}

Upon introducing a trigonometric deformation given by
\begin{equation}
\phi_{\varphi} = \cos{(\varphi)} ~~\mbox{with} ~~  \phi({\varphi}) = \sin{(\varphi)},
\label{topo001A}
\end{equation}
and assuming the ($+$ sign) result from Eq.~(\ref{topo06}), one obtains
\begin{eqnarray}
x_{\varphi} &=& y_{\phi} \varphi_{\phi} = (1 - \phi^2) \frac{1}{\phi_{\varphi}} = \cos{(\varphi)} = \sech{(\x)},
\label{topo001B}
\end{eqnarray}
with
\begin{eqnarray}
U(\varphi) &=& \frac{1}{2} x_{\varphi}^2 = \frac{1}{2}\cos{(\varphi)}^2 = \frac{1}{2}\sech{(\x)}^2,
\label{topo001C}
\end{eqnarray}
from which one can easily depict the localized energy density as
\begin{equation}
\rho(\varphi(\x)) = x_{\varphi}^2 = \sech{(\x)}^2.
\end{equation}
This solution corresponds to a simplified version of the {\em sine-Gordon} potential, for which the asymptotic behavior and the conserved topological quantities are easily obtained.

The potential $U(\varphi)$ engenders a set of critical points, $\varphi^{0}_{(s)}$, such that
\begin{equation}
U(\varphi^{0}_{(s)}) = 0 ~~ \mbox{and} ~~ \left.\frac{d U}{d\varphi}\right|_{\varphi = \varphi^{0}_{(s)}} = 0.
\end{equation}
By identifying $\varphi({\phi})$ with
\begin{equation}
\varphi({\phi}) = \arcsin{[\phi]} = \arcsin{[\tanh{(\x)}]},
\end{equation}
one notices that $\varphi^{0}_{(s)}$ corresponds to the asymptotic values of the kink solution, i. e.
\begin{equation}
\varphi^{0}_{(s)} = \varphi(\x\rightarrow\pm\infty) = (2s \pm 1)\frac{\pi}{2}, ~~ s \,\in\,\mathbb{Z}.
\end{equation}
The maximal points of $U(\varphi)$ are correspondingly provided by
\begin{equation}
\varphi^{Max}_{(s)} = \varphi(\x\rightarrow 0) = s\, \pi, ~~ s \,\in\,\mathbb{Z},
\end{equation}
and the topological charge is computed through
\begin{equation}
Q^{\varphi} = \left|\int^{+\infty}_{-\infty}{d\x \, x_{\varphi}}\right|
= \left|\varphi^{0}_{+} - \varphi^{0}_{-}\right| = \pi.
\end{equation}
The total energy of the localized solution is given by
\begin{equation}
M^{\varphi} = \left|\int^{+\infty}_{-\infty}{d\x \, x_{\varphi}^2}\right|
= 2.
\end{equation}

Taking some advantage of knowing the analytical behavior of $\lambda \phi^4$ solutions, the $3$-cyclic deformation chain is accomplished by an {\em ansatz} function described by
\begin{equation}
\psi^{(n)}(\phi) = y_{\phi}^{1/n} = (1 - \phi^2)^{\frac{1}{n}} = \sech{(\x)}^{\frac{2}{n}},
\end{equation}
that leads to
\begin{equation}
\psi^{(n)}_{\phi} = - \frac{2}{n} \phi (1 - \phi^2)^{\frac{1-n}{n}} =
- \frac{2}{n} \tanh{(\x)}\sech{(\x)}^{\frac{2(1-n)}{n}},
\end{equation}
such that
\begin{eqnarray}
z_{\psi} &=& y_{\phi} \psi_{\phi} = -\frac{2}{n} \psi \left(1 - \psi^n\right)^{\frac{1}{2}} =
- \frac{2}{n} \tanh{(\x)}\sech{(\x)}^{\frac{2}{n}},
\label{topo001B2}
\end{eqnarray}
with
\begin{eqnarray}
W(\psi) &=& \frac{1}{2} z_{\psi}^2 = \frac{2}{n^2}\psi^2 \left(1 - \psi^n\right) =
\frac{2}{n^2} \tanh{(\x)}^2\sech{(\x)}^{\frac{4}{n}},
\label{topo001C2}
\end{eqnarray}
from which the localized energy density is obtained as
\begin{equation}
\rho(\psi(\x)) =  z_{\psi}^2 = \frac{4}{n^2} \tanh{(\x)}^2\sech{(\x)}^{\frac{4}{n}}.
\end{equation}
This solution corresponds to the first family of models involving non-topological ($Q^{\psi} = 0$) {\em bell}-shaped solutions obtained in \cite{Bas03}.
{\em Bell}-shaped lump-like defects are non-topological structures generated by nonlinear interactions present at real scalar field models in ($1 ,\, 1$) spacetime dimensions \cite{Bas03}.

The potential $W(\psi)$ engenders one minimal point, $\psi^{0} = 0$, and the boundary points $\psi^{(n)0} = \, (\pm 1)^{(n+1)}$, for which
\begin{equation}
W(\psi^{(n)}) = 0,
\end{equation}
i. e. it has two {\em zeros} for $n$ odd (or $n < 1$) and three {\em zeros} for $n$ even.
Since we have a lump-like solution, the minimal point $\psi^{0} = 0$ is analytically connected to the asymptotic values of the $\lambda \phi^4$ kink solution previously obtained.
Through the same way, $W(\psi)$ has its maximal points corresponding to
\begin{equation}
\psi^{(n)Max}= (\pm 1)^{(n+1)} \left(\frac{2}{2+n}\right)^{\frac{1}{n}}.
\end{equation}
Finally, the total energy of the localized solution is given by
\begin{equation}
M^{\psi}_{(n)} = \left|\int^{+\infty}_{-\infty}{d\x \, z_{\psi}^2}\right| =
2 \frac{\sqrt{\pi}}{n^2} \frac{\Gamma\left( \frac{2}{n}\right)}{\Gamma\left(\frac{3}{2} + \frac{2}{n}\right)},
\end{equation}
from which one has, for instance,
\begin{equation}
M^{\psi}_{(1/2)} = \frac{512}{315},~~
M^{\psi}_{(1)} = \frac{16}{15},~~
M^{\psi}_{(2)} = 2,~~\mbox{and}~~
M^{\psi}_{(4)} = \frac{\pi}{8}.~~
\end{equation}

For the sake of completeness, it is also interesting to notice that the analytical chain described by Eq.~(\ref{topo06}) leads to
\begin{equation}
\psi^{(n)}_{\varphi} = - \frac{2}{n} \sin{(\varphi)} \cos{(\varphi)}^{\frac{2-n}{n}},
\end{equation}
from which one obtains
\begin{equation}
\psi^{(n)}(\varphi) = \cos{(\varphi)}^{\frac{2}{n}}.
\end{equation}
One could also notice that, in case of $n=2$, a constraint relation involving the values of the topological masses can be established as
\begin{equation}
M^{\phi} + \kappa^2 \left(M^{\varphi} +M^{\psi}_{(2)}\right) = \frac{8 + 3 \kappa^2}{6},
\end{equation}
with arbitrary $\kappa$, which introduces a pertinent issue about the possibility of finding some systematic way for constraining topological masses of cyclically deformed defects.

The first column of Fig.~\ref{Fig01A} summarizes the properties of the above discussed solutions, by introducing an illustrative view of their analytical properties, which will be extended in the following subsection.

\subsection{Second case}

Let us now investigate a novel $3$-cyclic chain which, as the previous one, is based on a similar {\em ansatz} involving some analytical dependence of $\psi$ on $y_{\phi}$ of the {\em dimensionless} $\lambda \phi^4$ theory.
At this point, however, we shall abbreviate some discussions which escape from the scope of this preliminary section.

Upon choosing an {\em ansatz} for a deformation function given by
\begin{equation}
\psi^{(n)}(\phi) = \frac{1}{n} \ln{[1 + y_{\phi}]} = \frac{1}{n} \ln{[2 - \phi^2]} = \frac{1}{n} \ln{[1 + \sech{(\x)}^2]},
\end{equation}
one obtains
\begin{equation}
\psi^{(n)}_{\phi} = \frac{1}{n} \frac{y_{\phi\phi}}{1 + y_{\phi}} = -\frac{2}{n} \frac{\phi}{(2 - \phi^2)} =
-\frac{2}{n} \frac{\tanh{(x)}}{[1 + \sech{(\x)}^2]},
\end{equation}
that can be parameterized by
\begin{equation}
\psi^{(n)}_{\phi} =
\frac{2}{n} \exp{(-n\,\psi)}\left[2 - \exp{(n\,\psi)}\right]^{\frac{1}{2}},
\end{equation}
where again we are assuming the ($+$ sign) result from Eq.~(\ref{topo06}).

The $3$-cyclic deformation chain from Eq.~(\ref{topo03}) can thus be implemented through the following definitions for $\phi_{\varphi}$ and $\varphi_{\psi}$,
\begin{equation}
\phi_{\varphi} = \frac{2 - \phi^2}{2 + \phi^2},
\end{equation}
that leads to
\begin{equation}
\varphi(\phi) = 2 \sqrt{2} \arctanh{\left(\frac{\phi}{\sqrt{2}}\right)} - \phi
~~~~~\leftrightarrows~~~~~
\phi(\varphi) = [\varphi]^{-1}(\phi),
\label{sdssd}
\end{equation}
and
\begin{equation}
\varphi_{\psi} = -\frac{n\,(2 + \phi^2)}{2\,\phi},
\end{equation}
that leads to
\begin{equation}
\varphi(\psi) =  2 \sqrt{2} \arctanh\left[\left(1 - \frac{\exp{(n \,\psi)}}{2}\right)^{\frac{1}{2}}\right]
~~\leftrightarrows~~
\psi(\varphi) = \frac{1}{n} \ln{\left\{ 2 \left[ 1 - \tanh{\left(\frac{\varphi}{2\sqrt{2}}\right)}^2\right]\right\}}.
\end{equation}

The BPS states can be obtained through
\begin{eqnarray}
z_{\psi} &=& y_{\phi} \psi_{\phi}
= - \frac{2}{n} \left[1 - \exp{(-n\,\psi)}\right]\left[2 - \exp{(n\,\psi)}\right]^{\frac{1}{2}}
= - \frac{2}{n} \frac{\tanh{(\x)}\sech{(\x)}^{2}}{1 + \sech{(\x)}^{2}},
\end{eqnarray}
and
\begin{eqnarray}
x_{\varphi} &=& y_{\phi} \varphi_{\phi} = \left[1 - \exp{(-n\,\psi)}\right]\left[4 - \exp{(n\,\psi)}\right]
= 5 - 2\left(\frac{1}{g(\varphi)} + g(\varphi)\right)\nonumber\\
&=& \frac{(2 + \tanh{(\x)}^2)\sech{(\x)}^{2}}{1 + \sech{(\x)}^{2}},
\end{eqnarray}
with $g(\varphi)= 1 - \tanh{[\varphi/(2\sqrt{2})]}^2$, so that the respectively associated potentials shall follow the definitions from Eq.~(\ref{topo02B}).
Notice that the explicit dependence of $x_{\varphi}$ on $\varphi$ is given in terms of $\psi(\varphi)$ since $\phi(\varphi) = [\varphi]^{-1}(\phi)$ is not analytically invertible.

\begin{figure}
\vspace{-2.5 cm}
\centerline{\psfig{file= 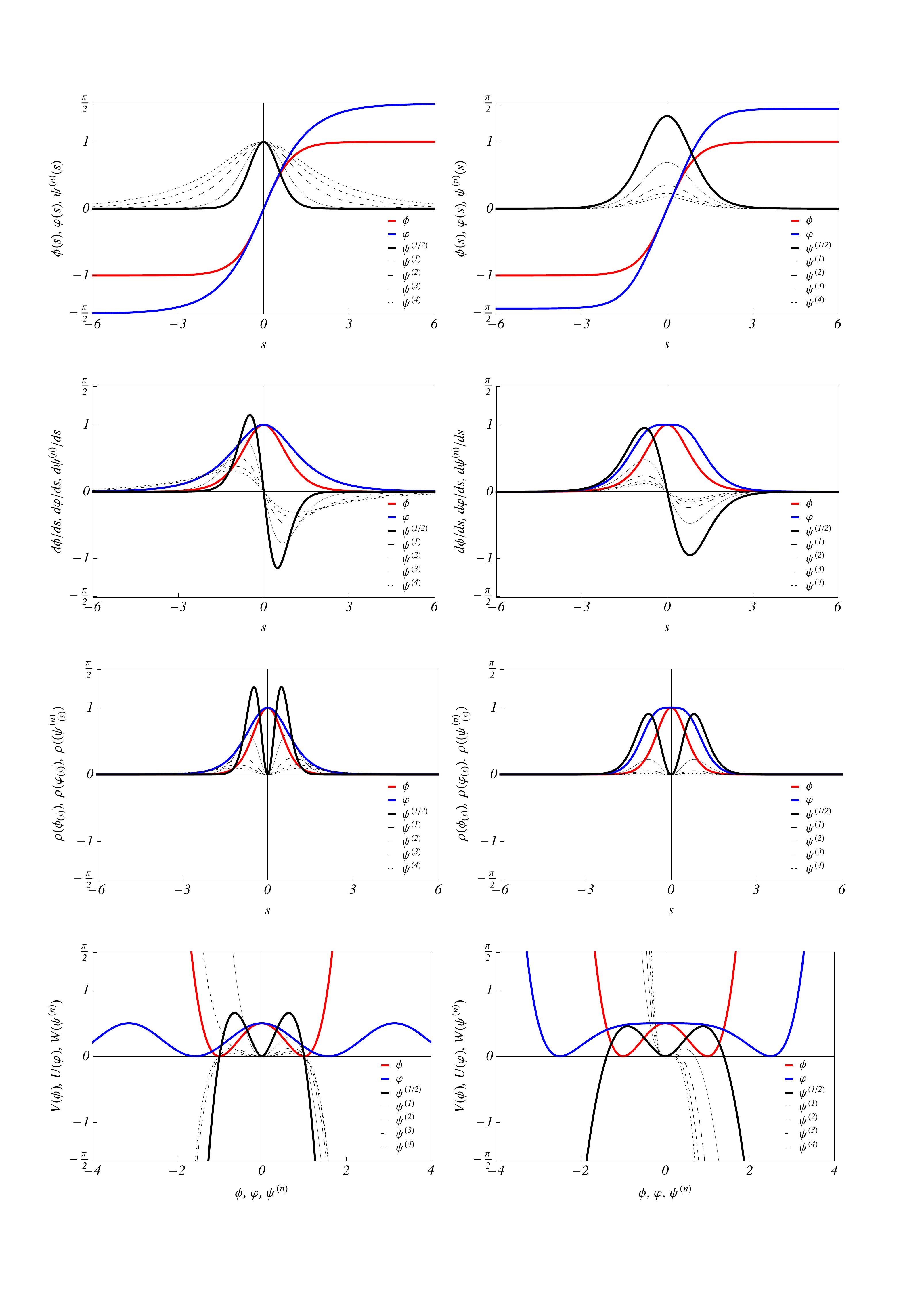, width= 17 cm}}
\vspace{-1.0 cm}
\caption{\footnotesize  First family (first column) and second family (second column) of $3$-cyclically deformed defects.
The figure shows the results for the primitive defects (red lines), $\phi(\x)$, $y_{\phi}(\x) = d\phi/d\x$, and $V(\phi)$, for the kink-like deformed defects (blue lines), $\varphi(\x)$, $x_{\varphi}(\x) = d\varphi/d\x$, and $U(\varphi)$, and for the lump-like deformed defects (black lines), $\psi^{(n)}(\x)$, $z_{\psi}(\x) = d\psi^{(n)}/d\x$, and $W(\psi)$.
The solutions related to $\psi^{(n)}(\x)$ are computed for $n$ = $1/2$ (thick line), $1$(thin line), $2$ (long-dashed line), $3$ (short-dashed line) and $4$ (dotted line).}
\label{Fig01A}
\end{figure}

The topological charge for the kink-like solution $\varphi(x)$ is given by
\begin{equation}
Q^{\varphi} = |\varphi^{0}_{+} - \varphi^{0}_{-}| = 4\sqrt{2} \arctanh{\left(\frac{1}{\sqrt{2}}\right)},
\end{equation}
since it is obtained from a potential $U(\varphi)$ that engenders two critical points, $\varphi^{0}_{\pm}$,
\begin{equation}
\varphi^{0}_{\pm} = \varphi(\x\rightarrow\pm\infty) = \pm 2\sqrt{2} \arctanh{\left(\frac{1}{\sqrt{2}}\right)},
\end{equation}
which also represent function {\em zeros} and correspond to the asymptotic values of the primitive $\lambda \phi^4$ kink solution.
The maximal point of $U(\varphi)$ is correspondingly given by
\begin{equation}
\varphi^{Max} = \varphi(\x\rightarrow 0) = 0,
\end{equation}
and is related  to the center of an analytical {\em plateau} symmetrically shaped around the origin, as one can depict from the second column of Fig.~(\ref{Fig01A}).

Otherwise, the non-topological lump-like solution $\psi^{(n)}(x)$ is obtained from a potential $W(\psi)$ that engenders one minimal point,  $\psi^{0} = 0$, and one boundary point $\psi^{(n)0} =  \frac{1}{n} \ln{(2)}$ which can be duplicated to its symmetrical partner, $\psi^{(n)0} =  - \frac{1}{n} \ln{(2)}$, if $1/n$ is an even integer for $n < 1$.
Therefore, it can have two or three {\em zeros}, depending on the values assumed by $n$.
Again, since we have {\em bell}-shaped solutions, the minimal points at $\psi^{0} = 0$ correspond to the asymptotic values of the $\lambda \phi^4$ kink solution previously obtained.
The potential $W(\psi)$ also has maximal points corresponding to $\psi^{(n)Max}= (1/n) \ln{[(\sqrt{17} -1)/2]}$.

Finally, the total energy of the localized solution is given by
\begin{equation}
M^{\varphi} = \left|\int^{+\infty}_{-\infty}{d\x \, x_{\varphi}^2}\right|
=\left[\frac{32}{3} - 10 \sqrt{2} \ln{(1+\sqrt{2})}\right],
\end{equation}
for the kink-like solution, and by
\begin{equation}
M^{\psi}_{(n)} = \left|\int^{+\infty}_{-\infty}{d\x \, z_{\psi}^2}\right| =
\frac{1}{n^2} \left[12 - 10 \sqrt{2} \ln{(1+\sqrt{2})}\right],
\end{equation}
for the lump-like solution, from which the equation constraining the values of topological masses can be established as
\begin{equation}
M^{\varphi} - M^{\phi} - n^2 M^{\psi}_{(n)} = 0.
\end{equation}

Fig.~\ref{Fig01A} summarizes the properties of the first (first column) and second (second column) families of non-systematically obtained cyclic deformations that we have discussed above.
For the first family, in the context of $3$-cyclic deformations, we have simply recovered the results for {\em sine-Gordon} and {\em bell}-shaped defects scrutinized by Bazeia {\it et al.} \cite{Bas01,Bas02,Bas03}.
It satisfactorily works as preliminary analysis.
The second family brings up novel results.
One can depict, for instance, the {\em plateau}-shaped form of the potential for the modified kink solution (blue lines), which denotes the possibility of a kind of slow-rolling behavior for the scalar field $\varphi$ when its dynamical properties are considered.
The plots show the results for the primitive defects (thick red lines), $\phi(\x)$, $y_{\phi}(\x) = d\phi/d\x$, and $V(\phi)$; for the kink-like deformed defects (thick blue lines), $\varphi(\x)$, $x_{\varphi}(\x) = d\varphi/d\x$, and $U(\varphi)$; and for the lump-like deformed defects (black lines), $\psi^{(n)}(\x)$, $z_{\psi}(\x) = d\psi^{(n)}/d\x$, and $W(\psi)$.
In particular, for solutions related to $\psi^{(n)}(\x)$ we have set $n = 1/2,\, 1,\, 2,\, 3$ and $4$ in order to verify all the properties related to their critical points and function {\em zeros} discussed above.

\begin{figure}
\centerline{\psfig{file= 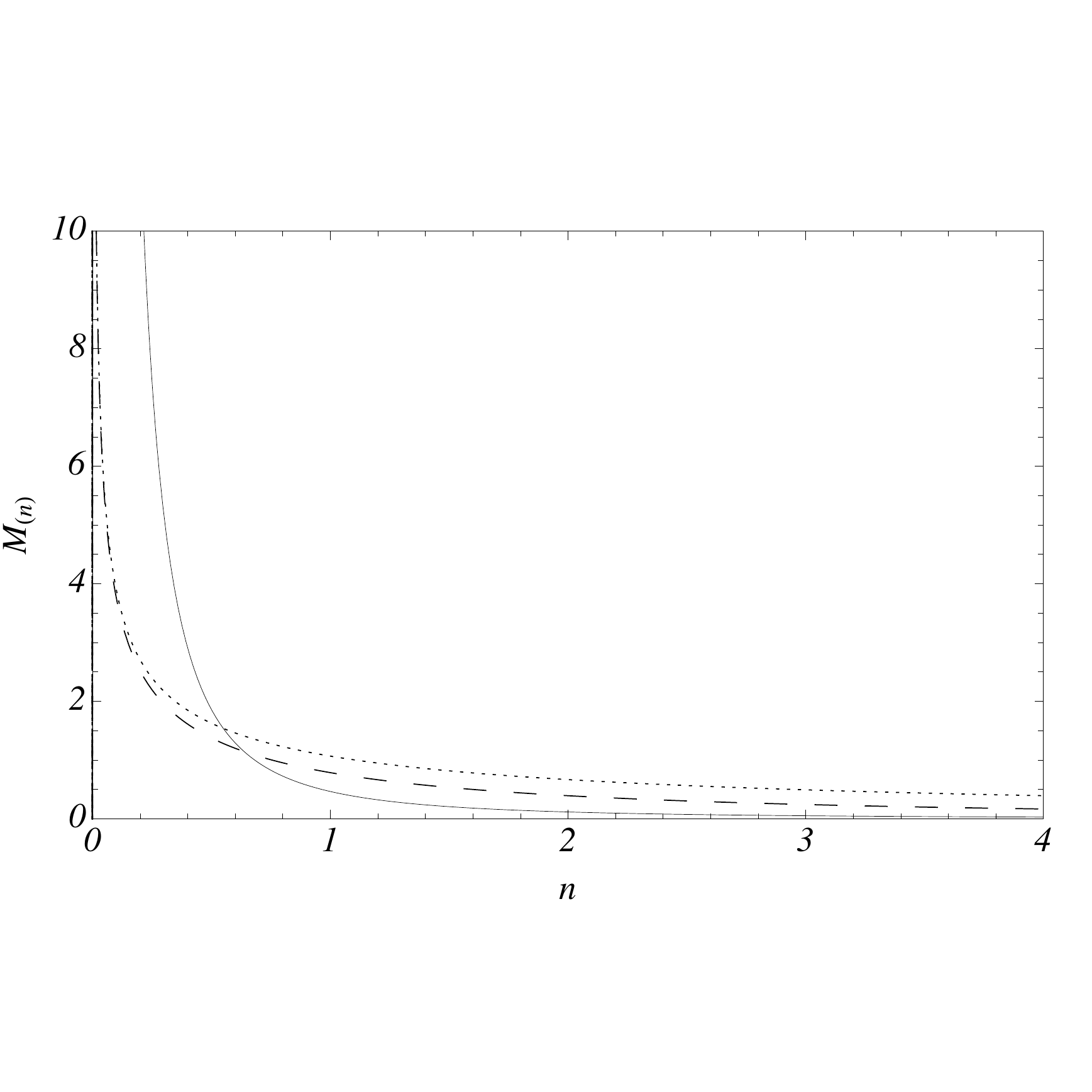, width= 14 cm}}
\caption{\small  Total energy of localized solutions (topological masses) as function of the parameter $n$.
The results for $M^{\psi}_{(n)}$ for the first family (dotted line), obtained from $\psi^{(n)}(\phi) = y_{\phi}^{(1/n)}$, and for the second family (solid line), obtained from $\psi^{(n)}(\phi) = \ln{(1 + y_{\phi})^{(1/n)}}$, are compared with preliminary results from \cite{Bas03} (dashed line).}
\label{Fig01C}
\end{figure}

In Fig.~\ref{Fig01C} we have merely illustrated the explicit analytical dependencies of the topological masses, $M^{\psi}_{(n)}$ on $n$, and compared one each other by including a third case investigated at \cite{Bas03}, for which the topological mass is given by
\begin{equation}
M_{(n)} = 2 \frac{\sqrt{\pi}}{n^2} \frac{\Gamma\left(\frac{1}{2} + \frac{2}{n}\right)}{\Gamma\left(2 + \frac{2}{n}\right)}.
\end{equation}
It is obtained from the BPS framework for which
\begin{eqnarray}
z_{\psi} &=&  - \frac{2}{n} \psi^{\frac{n}{2}+1}\left(1 -  \psi^{n}\right)^{\frac{1}{2}},
\end{eqnarray}
with
\begin{eqnarray}
\psi^{(n)}(\x) &=& (1 + \x^{2})^{-\frac{1}{n}}.
\end{eqnarray}

Just to summarize this introductory section, an interesting property related to the effective applicability of cyclic deformations can be depicted from Eq.~(\ref{sdssd}).
Although one can obtain the explicit analytical dependence of $\phi$ on $\varphi$, the corresponding analytical expression for the inverse function $\varphi(\phi) = [\phi]^{-1}(\varphi)$ cannot be obtained.
The problem is circumvented by the cyclic chain that relates $\phi \rightleftarrows \varphi \rightleftarrows \psi \rightleftarrows \phi$, and therefore allows one to obtain the explicit dependence of $x_{\varphi}$ on $\x$ and of $U(\varphi)$ on $\varphi$.

Finally, for all the above obtained solutions, the quantum mechanics correspondence can be directly obtained.
The related problem described by those {\em kind of} modified P\"{o}schl-Teller potentials computed from the second derivatives of the deformed potentials ($U$, $V$ and $W$) naturally supports normalized zero-mode solutions $\eta_{0}(\x)$ (respectively given by $x_{\varphi}(\x)$, $y_{\phi}(\x)$ and $z_{\psi}(\x)$) at zero energy.
In spite of being pertinent, the determination of additional bound state solutions and of the eventual ground state for solutions with multiple nodes (in case of lump-like related solutions) is out of the scope of this preliminary analysis.

\section{Systematic procedure for $3$-cyclic deformations}

The previous section has introduced the issue concerning
 the possibility of generating cyclically deformed defect structures through some systematic procedure.
Hereon we shall introduce of set of systematic rules based on asymptotically bounded trigonometric and hyperbolic deformation functions for obtaining constraint relations involving the topological masses of defects belonging to a cyclic chain.

One can find a large variety of defect structures involving hyperbolic and trigonometric deformation chains.
Our analysis, however, is concerned with the analytical expressions for deformation functions which guarantee the existence of closure relations constraining the topological masses of the deformed defects.
The BPS solutions and their respective localized energy distributions shall be analytically obtained for $3$-cyclic chains triggered by kink-like solutions of a conveniently renamed ({\em dimensionless}) $\lambda \chi^4$ theory.

Let us begin by replacing the scalar field $\varphi$ introduced in the previous analysis by a renamed scalar field, $\chi$, which, from this point, will designate the primitive kink defect that triggers the $3$-cyclic deformation chain.
All the equalities involving $\varphi$ and $x_{\varphi}$ into Eqs.(\ref{topo02}) are therefore replaced by
\begin{eqnarray}
w_{\chi}    &=& \frac{d \chi}{d \x} = y_{\phi} \chi_{\phi} = z_{\psi} \chi_{\psi},
\label{topo023}
\end{eqnarray}
such that
\begin{eqnarray}
y_{\phi}    &=& w_{\chi} \phi_{\chi}, \nonumber\\
z_{\psi}    &=& w_{\chi} \psi_{\chi},
\label{topo023B}
\end{eqnarray}
and the cyclically derived BPS potentials belonging to the $3$-cyclic deformation chain are now given by
\begin{equation}
T(\chi) = \frac{1}{2} w_{\chi}^2 ~\rightleftarrows ~
V(\phi) = \frac{1}{2} y_{\phi}^2 ~\rightleftarrows ~
W(\psi) = \frac{1}{2} z_{\psi}^2 ~\rightleftarrows ~
T(\chi) = \frac{1}{2} w_{\chi}^2,
\label{topo023C}
\end{equation}
In this context, the novel deformation functions shall be constrained by the chain rule given by
\begin{equation}
\psi_{\phi} \phi_{\chi} \chi_{\psi}= 1.
\label{topo033}
\end{equation}

\subsection{Hyperbolic Deformation}

Let us consider the set of auxiliary derivatives described by
\begin{eqnarray}
\psi^{(n)}_{\chi} &=& \tanh{(n\,\chi)}, \nonumber\\
\phi^{(n)}_{\chi} &=& \sech{(n\,\chi)},
\label{hyp03}
\end{eqnarray}
which upon straightforward integrations lead to
\begin{eqnarray}
\psi^{(n)}(\chi)    &=& -\frac{1}{n}\ln{\left[\frac{\cosh{(n\,\chi)}}{\cosh{(n)}}\right]}, \nonumber\\
\phi^{(n)}(\chi) &=& \frac{2}{n}\arctan{\left[\tanh{\left(\frac{n\,\chi}{2}\right)}\right]},
\label{hyp3}
\end{eqnarray}
with constants chosen to fit ordinary values for the asymptotic limits.
After simple mathematical manipulations involving the hyperbolic fundamental relation,
\begin{eqnarray}
\tanh{(n\,\chi)}^2 + \sech{(n\,\chi)}^2 &=& 1,
\label{hypA3}
\end{eqnarray}
with relations from Eq.~(\ref{topo023B}), one easily identifies the following closure relation,
\begin{eqnarray}
w_{\chi}^{2}
&=& w_{\chi}^{2}\left[\tanh{(n\,\chi)}^2 + \sech{(n\,\chi)}^2\right]\nonumber\\
&=& w_{\chi}^{2}\left[\psi^{(n)\,2}_{\chi} + \phi^{(n) \, 2}_{\chi}\right]\nonumber\\
&=& z_{\psi}^{2} + y_{\phi}^{2}
,\label{hyp013}
\end{eqnarray}
which constrains the localized energy distributions for $3$-cyclic deformed defect structures.

For the sake of completeness, one should have
\begin{eqnarray}
\chi &\equiv& \chi(\psi) = \frac{1}{n}\arccosh{\left[\cosh{(n)}\,\exp{(-n\, \psi)}\right]} \nonumber\\
     &\equiv& \chi(\phi) = \frac{2}{n}\arctanh{\left[\tan{(n\, \phi/2)}\right]},
\label{s133}
\end{eqnarray}
and, in case of assuming,
\begin{equation}
\chi(\x) = \tanh{(\x)} ~~ \mbox{with} ~~ w_{\chi} = 1 - \chi^2,
\label{s13}
\end{equation}
the BPS deformed functions would concomitantly obtained as
\begin{eqnarray}
y_{\phi}    &=& w_{\chi} \phi_{\chi} = (1 - \chi^2) \, \tanh{(n\,\chi)}, \nonumber\\
z_{\psi}    &=& w_{\chi} \psi_{\chi} = (1 - \chi^2) \, \sech{(n\,\chi)},
\label{hyp003}
\end{eqnarray}
with the respective dependencies on $\phi^{(n)}(\x)$ and $\psi^{(n)}(\x)$ implicitly given by Eq.~(\ref{s133}).

The first column of Fig.~\ref{Fig03A} shows the analytical dependence on $\x$ for the $3$-cyclically deformed defects obtained through the hyperbolic deformation functions from Eq.(\ref{hyp3}) .
For the primitive defect engendered by the $\lambda \chi^4$ theory (c. f. Eq.~(\ref{s13})), one has topological kink-like defects for $\phi^{(n)}$ and lump-like defects for $\psi^{(n)}$.
The plots show the results for the primitive defects, $\chi(\x)$, $w_{\chi}(\x) = d\chi/d\x$, and $\rho(\chi(\x))$; and for the corresponding deformed defects, $\phi^{(n)}(\x)$, $y_{\phi}(\x) = d\phi^{(n)}/d\x$, and $\rho(\phi^{(n)}(\x))$; and $\psi^{(n)}(\x)$, $z_{\psi}(\x) = d\psi^{(n)}/d\x$, and $\rho(\psi^{(n)}(\x))$.
We have set $n = k \pi /8$ with $k$ in the range between $1$ and $4$, in order to depict the analytical behavior driven by the free parameter $n$.

The corresponding BPS potentials, $W(\psi)$ and $V(\phi)$, for the same set of $n$ values, are depicted from the first column of Fig.~\ref{Fig03B}.
The potential $V(\phi)$ engenders a set of critical points, $\phi^{(n)0}_{\pm}$, which correspond to the asymptotic values of the kink-like solution, i. e.
\begin{equation}
\phi^{(n)0}_{\pm} = \phi^{(n)}(\x\rightarrow \pm \infty) = \pm\frac{2}{n} \arctan{\left[\tanh{\left(\frac{n}{2}\right)}\right]}
\end{equation}
such that the topological charge is given by
\begin{equation}
Q^{\phi}_{(n)} = \frac{4}{n} \arctan{\left[\tanh{\left(\frac{n}{2}\right)}\right]},~~\mbox{with} ~~ n > 0.
\label{kinkstopo}
\end{equation}
The topological masses are (semi)analytically obtained as function of the free parameter $n$ as
\begin{equation}
M^{\phi}_{(n)} = \frac{2}{n^2} \ln{\left[2 +  \exp{(2n)} + \exp{(-2n)}\right]} - \frac{1}{n^3}\left\{ Li_{2}\left[-e^{(-2n)}\right] - Li_{2}\left[-e^{(2n)}\right]\right\}
\end{equation}
where
\begin{equation}
Li_{s}(t) = \sum_{j=1}^{\infty}{\frac{t^j}{j^s}}
\end{equation}
is the Jonqui\`eres ({\em polylogarithm}) function.

The non-topological lump-like solution $\psi^{(n)}(x)$ is obtained from a potential $W(\psi)$ that engenders one minimal point,  $\psi^{(n)0} = 0$, and one boundary point $\psi^{(n)0} = \psi^{(n)} (\x\rightarrow\pm\infty) =  \frac{1}{n} \ln{[\cosh{(n)}]}$ which can be duplicated to its symmetrical partner, $\psi^{(n)0} \rightarrow -\psi^{(n)0}$, if $1/n$ is an even integer for $n < 1$.
In this case, the total energy of the localized solution is given by
\begin{equation}
M^{\psi}_{(n)} = \frac{4}{3} - \frac{2}{n^2} \ln{\left[2 +  \exp{(2n)} + \exp{(-2n)}\right]} + \frac{1}{n^3}\left\{ Li_{2}\left[-e^{(-2n)}\right] - Li_{2}\left[-e^{(2n)}\right]\right\}.
\end{equation}

\subsection{Trigonometric Deformation}

Let us now turn to a set of trigonometric auxiliary functions described by
\begin{eqnarray}
\psi^{(n)}_{\chi} &=& -\sin{(n\,\chi)}, \nonumber\\
\phi^{(n)}_{\chi} &=&  \cos{(n\,\chi)},
\label{hyp03B}
\end{eqnarray}
which upon straightforward integrations lead to
\begin{eqnarray}
\psi^{(n)}(\chi)  &=& \frac{1}{n} \left[\cos{(n\,\chi)}-\cos{(n)}\right], \nonumber\\
\phi^{(n)}(\chi)  &=& \frac{1}{n} \sin{(n\,\chi)},
\label{hyp3B}
\end{eqnarray}
with constants chosen to fit ordinary values for the asymptotic limits.
After simple mathematical manipulations involving the ordinary trigonometric fundamental relation,
\begin{eqnarray}
\sin{(n\,\chi)}^2 + \cos{(n\,\chi)}^2 &=& 1,
\label{hypA3B}
\end{eqnarray}
followed by relations from Eq.~(\ref{topo023B}), one recurrently identifies the closure relation,
\begin{eqnarray}
w_{\chi}^{2}
&=& w_{\chi}^{2}\left[\sin{(n\,\chi)}^2 + \cos{(n\,\chi)}^2\right]\nonumber\\
&=& w_{\chi}^{2}\left[\psi^{(n)\,2}_{\chi} + \phi^{(n)\,2}_{\chi}\right]\nonumber\\
&=& z_{\psi}^{2} + y_{\phi}^{2}.
\label{hyp013B}
\end{eqnarray}
Again, one should have
\begin{eqnarray}
\chi &\equiv& \chi(\psi) = \frac{1}{n}\arccos{\left[n\, \psi + \cos{(n)}\right]} \nonumber\\
     &\equiv& \chi(\phi) = \frac{1}{n}\arcsin{\left[n\, \phi\right]},
\label{s133B}
\end{eqnarray}
and, in case of assuming that $\chi(\x)$ is given by Eq.~(\ref{s13}), the BPS deformed functions are given by
\begin{eqnarray}
y_{\phi}    &=& w_{\chi} \phi_{\chi} =  (1 - \chi^2) \, \cos{(n\,\chi)}, \nonumber\\
z_{\psi}    &=& w_{\chi} \psi_{\chi} = -(1 - \chi^2) \, \sin{(n\,\chi)},
\label{hyp003B}
\end{eqnarray}
with the respective dependencies on $\phi^{(n)}(\x)$ and $\psi^{(n)}(\x)$ obtained from Eq.~(\ref{s133B}).

\begin{figure}
\centerline{\psfig{file= 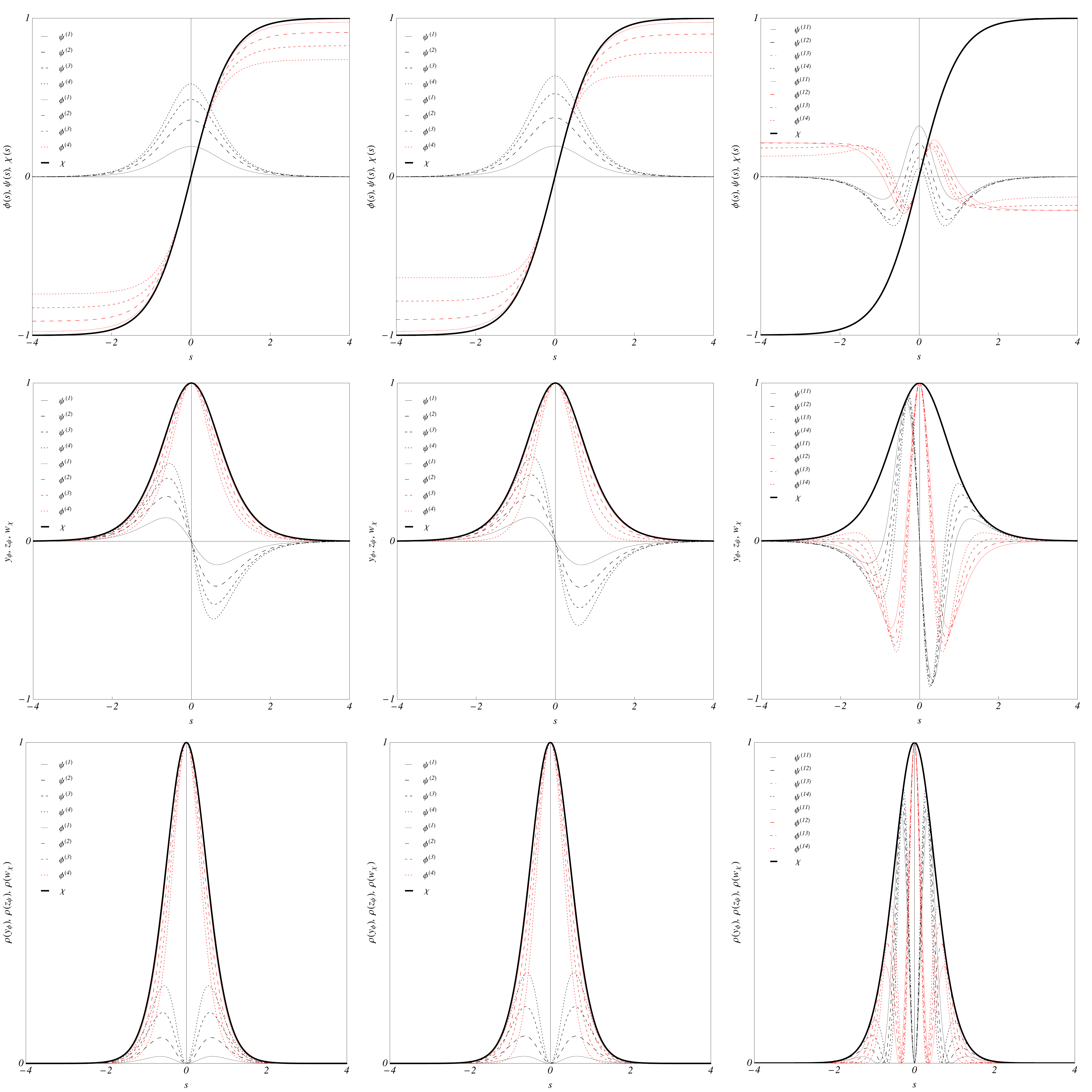,width= 18 cm}}
\caption{\small  The $3$-cyclically deformed defects obtained from hyperbolic (first column) and trigonometric (second and third columns) deformation functions.
Results are for the primitive $\lambda \chi^4$ kink solution, $\chi(\x)$ (thick black line), for the kink-like deformed defects, $\psi^{(n)}(\x)$ (black lines), and for the lump-like deformed defects, $\phi^{(n)}(\x)$ (red lines).
The free parameter $n$ was set equal to $k \pi /8$ with $k$ assuming the following integer values in the first/second (third) columns: $1(11)$ for solid lines, $2(12)$ for long-dashed lines, $3(13)$ for short-dashed lines, and $4(14)$ for dotted lines.}
\label{Fig03A}
\end{figure}

The second and third columns of Fig.~\ref{Fig03A} show the analytical dependence on $\x$ for the $3$-cyclically deformed defects obtained through the trigonometric deformation functions from Eq.(\ref{hyp3B}).
As in the hyperbolic case, the primitive defect from Eq.~(\ref{s13}) leads to kink-like defects for $\phi^{(n)}$ and lump-like defects for $\psi^{(n)}$.
We have set $n = k \pi /8$ with $k$ in the range between $1$ and $4$ in the second column and in the range between  $11$ and $14$ in the third column.
Large values for $k$ highlight the oscillatory properties introduced by trigonometric deformations.
By comparing the results from the first column (hyperbolic deformation) with those from the second column, one is able to verify  that for small values of $n$, the oscillatory pattern due to trigonometric deformations disappears, and hyperbolic and trigonometric deformations are close to each other.

\begin{figure}
\centerline{\psfig{file= 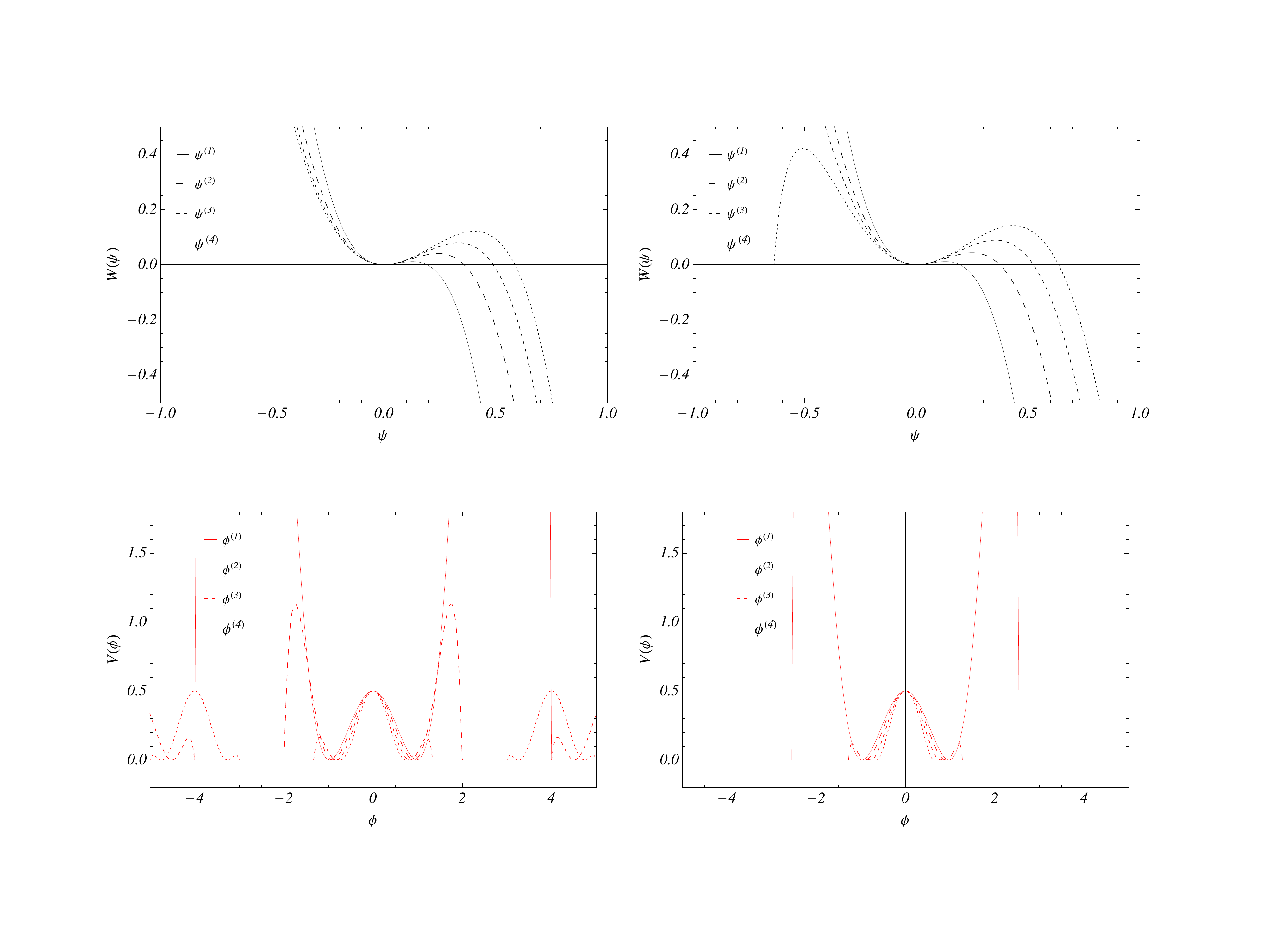,width= 22 cm}}
\caption{\small  The $3$-cyclically deformed potentials obtained from hyperbolic (first column) and trigonometric (second) deformation functions.
The legend parameters are in correspondence with those of first and second columns from Fig.~(\ref{Fig03A}).}
\label{Fig03B}
\end{figure}

The corresponding BPS potentials, $W(\psi)$ and $V(\phi)$, for the same set of $n$ parameters, are depicted from the second column of Fig.~\ref{Fig03B}.
The potential $V(\phi)$ engenders a set of critical points, $\phi^{(n)0}_{\pm}$, which correspond to the asymptotic values of the kink-like solution, i. e.
\begin{equation}
\phi^{(n)0}_{\pm} = \phi^{(n)}(\x\rightarrow \pm \infty) = \pm\frac{\sin{(n)}}{n}
\end{equation}
such that the topological charge and the topological mass are respectively given by
\begin{equation}
Q^{\phi}_{(n)} =
2\left|\frac{\sin{(n)}}{n}\right|
\label{kinkstopoB}
\end{equation}
and
\begin{equation}
M^{\phi}_{(n)} =\frac{2}{3} - \frac{2n \cos{(2n)} -\sin{(2n)}}{4n^3}.
\end{equation}

The non-topological lump-like solution $\psi^{(n)}(x)$ is obtained from a potential $W(\psi)$ that engenders one minimal point,  $\psi^{0} = 0$, and two function {\em zeros} given by
\begin{equation}
\psi^{(n)0}_{\pm} = \psi^{(n)} (\x\rightarrow\pm\infty) =  \frac{[\pm 1 - \cos{(n)]}}{n}.
\end{equation}
In this case, the corresponding total energy of the localized solution is given by
\begin{equation}
M^{\psi}_{(n)} = \frac{2}{3} + \frac{2n \cos{(2n)} -\sin{(2n)}}{4n^3}.
\end{equation}

In correspondence with Fig. \ref{Fig03A}, the plots of the BPS deformed potentials for hyperbolic (first column) and trigonometric (second column) deformation functions can be depicted from Fig.~\ref{Fig03B}.
We have suppressed the results for large values of $n$ (corresponding to the third column of Fig. \ref{Fig03A}) from Fig. \ref{Fig03B} since the critical points cannot be visually depicted from the plots, in spite of being calculable.

\begin{figure}
\centerline{\psfig{file= 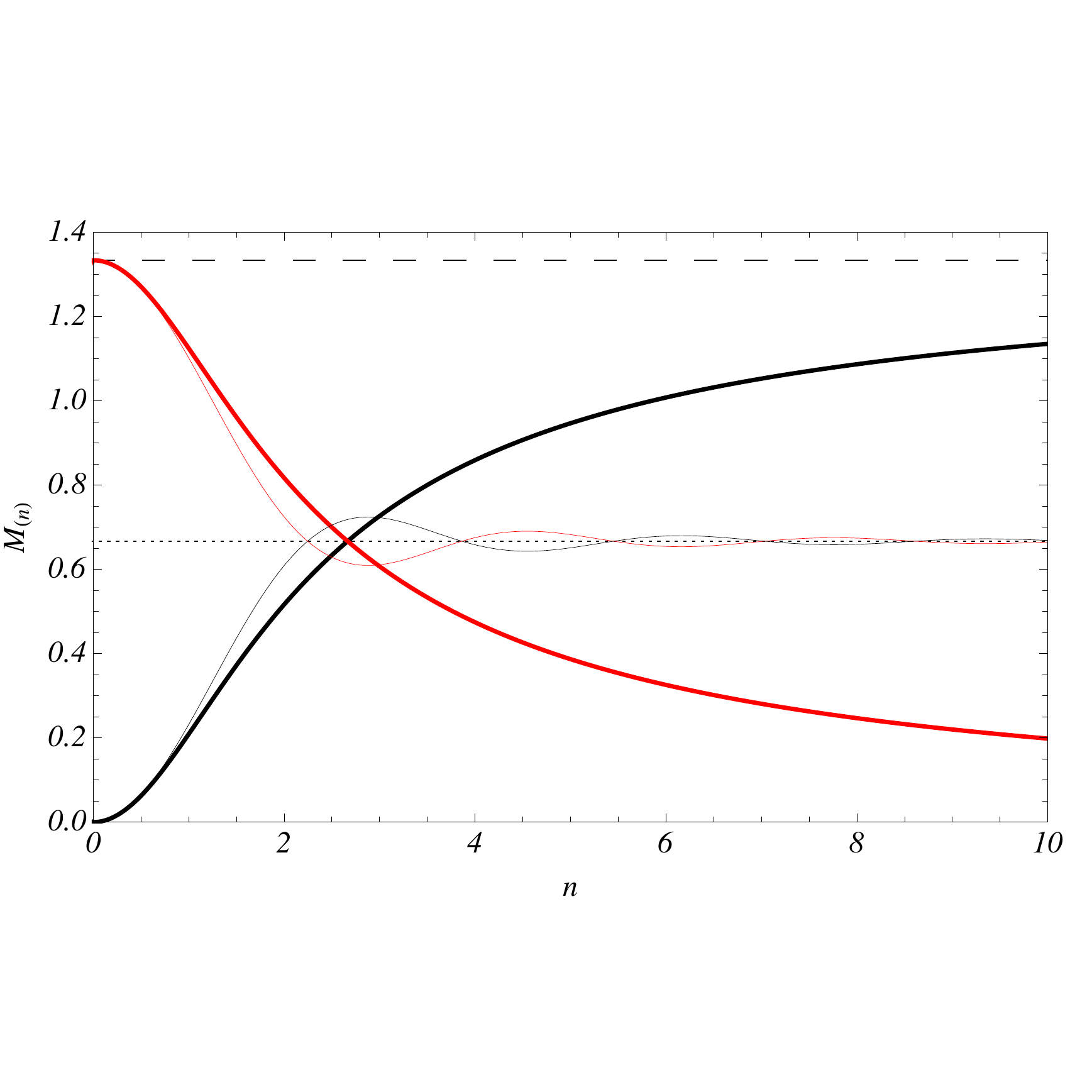,width= 14 cm}}
\caption{\small  Total energy of localized solutions (topological masses, $M_{(n)}$) as function of the parameter $n$ for hyperbolic (thick lines) and trigonometric (thin lines) $3$-cyclic deformation chains.
Results are for $M^{\phi}_{(n)}$ (red lines) and $M^{\psi}_{(n)}$ (black lines).}
\label{Fig03C}
\end{figure}

To partially end up, Fig.~\ref{Fig03C} resumes the results for the topological masses analytically obtained as function of the free parameter $n$, from which the constraint equation obtained from Eqs.~(\ref{hyp013}) and (\ref{hyp013B}), is ultimately verified for both hyperbolic and trigonometric $3$-cyclic deformation chains as
\begin{equation}
M^{\psi}_{(n)} + M^{\phi}_{(n)}  = M^{\chi} = \frac{4}{3}.
\end{equation}

\section{Systematic procedure for $4$-cyclic deformations}

From this point, we shall verify the possibility of extending the analysis performed in the previous section to a $4$-cyclic deformation chain, for which a modified constraint relation involving the topological masses is obtained.
The analytical results for topological charges and topological masses shall be straightforwardly obtained.
However, the BPS solutions and their respective localized energy distributions will be merely illustrated.
Likewise, the asymptotic limit analysis involving the computation of critical points and function {\em zeros} shall be suppressed since they can be performed as immediate extensions of the preliminary studies.

Let us hence recover the scalar field $\varphi$ which shall be reintroduced into a novel $4$-cyclic deformation chain.
The additional scalar field, $\chi$, continues to designate the primitive defect that triggers the $4$-cyclic deformations.
The coupled system described by Eqs.~(\ref{topo02}) shall thus be complemented by
\begin{eqnarray}
w_{\chi}    &=& \frac{d \chi}{d \x} = x_{\varphi} \chi_{\varphi} = y_{\phi} \chi_{\phi} = z_{\psi} \chi_{\psi},
\label{topo024}
\end{eqnarray}
such that
\begin{eqnarray}
x_{\varphi} &=& w_{\chi} \varphi_{\chi}, \nonumber\\
y_{\phi}    &=& w_{\chi} \phi_{\chi}, \nonumber\\
z_{\psi}    &=& w_{\chi} \psi_{\chi},
\label{topo024B}
\end{eqnarray}
and the cyclically derived BPS potentials belonging to the $4$-cyclic deformation chain will be given by
\begin{equation}
T(\chi) = \frac{1}{2} w_{\chi}^2 ~\rightleftarrows ~
U(\varphi) = \frac{1}{2} x_{\varphi}^2 ~\rightleftarrows ~
V(\phi) = \frac{1}{2} y_{\phi}^2 ~\rightleftarrows ~
W(\psi) = \frac{1}{2} z_{\psi}^2 ~\rightleftarrows ~
T(\chi) = \frac{1}{2} w_{\chi}^2,
\label{topo024C}
\end{equation}
Hyperbolic and trigonometric deformation functions shall naturally follow the chain rule given by
\begin{equation}
\psi_{\phi} \phi_{\varphi} \varphi_{\chi} \chi_{\psi}= 1.
\label{topo034}
\end{equation}

\subsection{Hyperbolic Deformation}

Let us consider the set of auxiliary functions described by
\begin{eqnarray}
\psi^{(n)}_{\chi} &=& \tanh{(n\,\chi)}, \nonumber\\
\phi^{(n)}_{\chi} &=& \tanh{(n\,\chi)}\sech{(n\,\chi)}, \nonumber\\
\varphi^{(n)}_{\chi} &=& \sech{(n\,\chi)}^{2},
\label{hyp00}
\end{eqnarray}
which, for a given  $w_{\chi}$ substituted into Eq.(\ref{topo024B}), completes the $4$-cyclic chain.
Upon straightforward integrations the above deformation functions lead to
\begin{eqnarray}
\psi^{(n)}(\chi)     &=& -\frac{1}{n}\ln{\left[\frac{\cosh{(n\,\chi)}}{\cosh{(n)}}\right]}, \nonumber\\
\phi^{(n)}(\chi)     &=& \frac{1}{n}\left[\sech{(n\,\chi)} -\sech{(n)} \right], \nonumber\\
\varphi^{(n)}(\chi) &=& \frac{1}{n}\tanh{(n\,\chi)},
\label{hyp}
\end{eqnarray}
with constants chosen to fit ordinary asymptotic values.
By following the same simple mathematical manipulations involving the fundamental relation from Eq.(\ref{hypA3}), one easily identifies that
\begin{eqnarray}
w_{\chi}^{2}
&=& w_{\chi}^{2}\left[\tanh{(n\,\chi)}^2 + \sech{(n\,\chi)}^2\left(\tanh{(n\,\chi)}^2 + \sech{(n\,\chi)}^2\right)\right]\nonumber\\
&=& w_{\chi}^{2}\left[\psi^{(n)\,2}_{\chi} + \phi^{(n)\,2}_{\chi} + \varphi^{(n)\,2}_{\chi}\right]\nonumber\\
&=& z_{\psi}^{2} + y_{\phi}^{2} + x_{\varphi}^{2}.
\label{hyp1}
\end{eqnarray}
It certainly constrains the values of the topological masses obtained from hyperbolically deformed defects.

\begin{figure}
\centerline{\psfig{file= 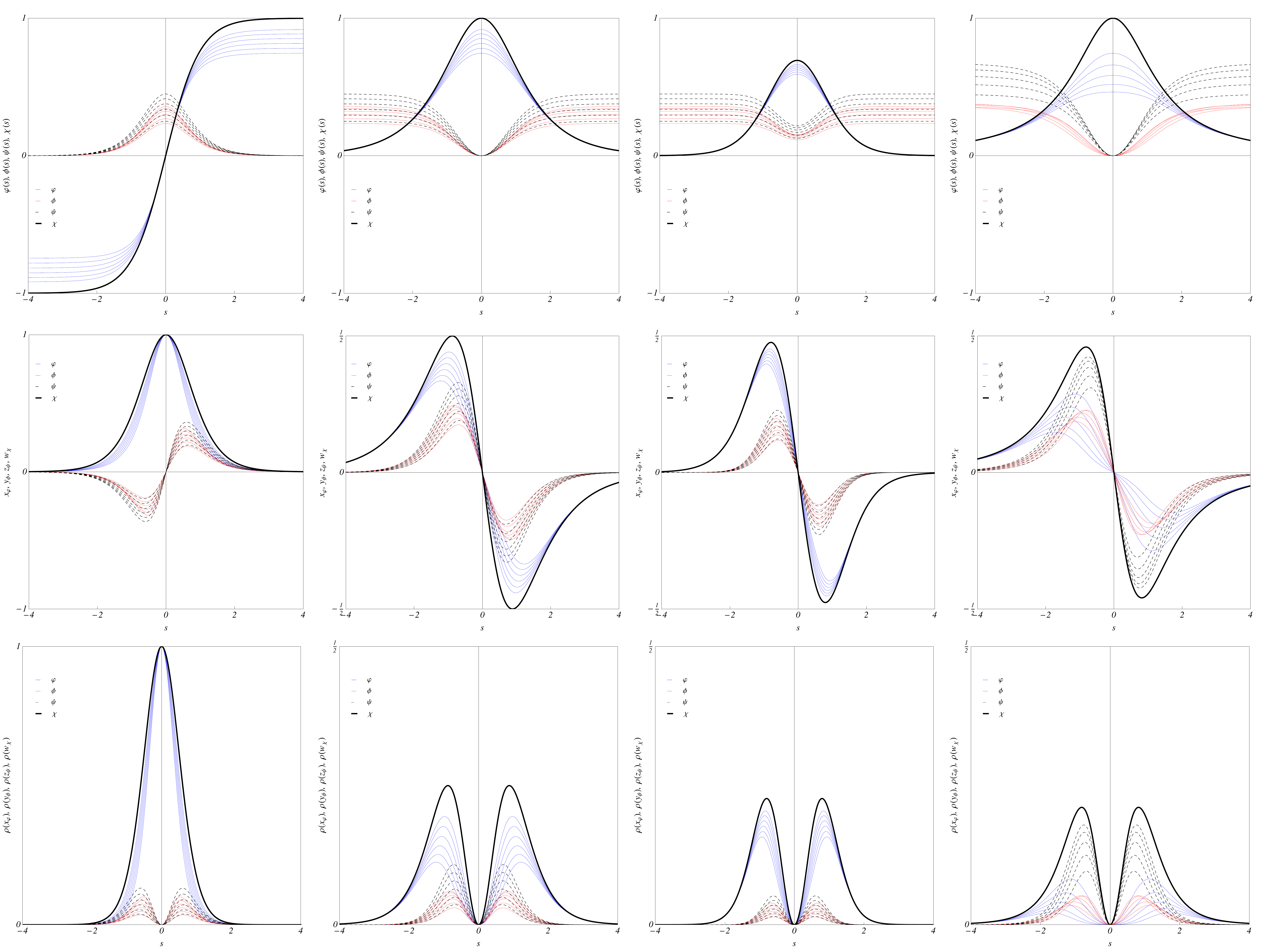,width= 18 cm}}
\caption{\small  The $4$-cyclically deformed defects obtained from {\em hyperbolic} deformation functions.
Results from each column are for four different triggering solutions for $\chi(\x)$, in respective correspondence with Eqs.(\ref{s1}-\ref{s4}).
In the first column results are for the primitive $\lambda \chi^4$ kink solution, $\chi(\x)$ (thick black line), which triggers the $4$-cyclic chain with the kink-like deformed defects, $\varphi^{(n)}(\x)$ (blue lines), a first family of lump-like deformed defects, $\phi^{(n)}(\x)$ (red lines), and a second family of lump-like deformed defects $\phi^{(n)}(\x)$ (thin black lines).
We have set $n = k \pi /6$ with $k$ in the range between $1$ and $2$, by steps of $0.2$ units, in order to qualitatively depict the analytical profile related to $n$.
Second, third and forth columns are merely illustrative.}
\label{Fig04A}
\end{figure}

From this point, the explicit dependence on $\x$ is directly obtained from $\chi(\x)$.
Four different forms of introducing the triggering primitive defect $\chi$ can be pictorially considered: the $\lambda \chi^4$ kink solution,
\begin{equation}
\chi_1(\x) = \tanh{(\x)} ~~ (\mbox{in the first column}),
\label{s1}
\end{equation}
a lump-like $\lambda \chi^4$ deformed solution,
\begin{equation}
\chi_2(\x) = \sech{(\x)} ~~ (\mbox{in the second column}),
\label{s2}
\end{equation}
a lump-like $\lambda \chi^4$ logarithmically deformed solution (engendered by the {\em plateau}-shaped potential from section II),
\begin{equation}
\chi_3(\x) = \ln{\left[1 + \sech{(\x)}^2\right]} ~~ (\mbox{in the third column}),
\label{s3}
\end{equation}
and a {\em bell}-shaped $\lambda \chi^4$ deformed solution,
\begin{equation}
\chi_4(\x) = 2 \left[2 + \x^2\right]^{-1} ~~ (\mbox{in the forth column}),
\label{s4}
\end{equation}
where the last three cases are merely illustrative, in the sense that we have not obtained the corresponding analytical expressions for their topological masses, and the information between {\em parenthesis} are in correspondence with Figs.~\ref{Fig04A} and ~\ref{Fig04B}.
The BPS deformed functions obviously depend on the choice of $\chi(\x)$.

Fig.~\ref{Fig04A} shows the analytical dependence on $\x$ for the $4$-cyclically deformed defects obtained through the hyperbolic deformation functions from Eq.(\ref{hyp00}).
For the primitive defect engendered by the {\em dimensionless} $\lambda \chi^4$ theory (c. f. Eq.~(\ref{s1})), one has topological kink-like defects solely for $\varphi^{(n)}$ given by
\begin{equation}
\varphi^{(n)}(\x) = \sech{(n\,\tanh{(\x)})}^{2},
\label{kink}
\end{equation}
where
\begin{equation}
x_{\varphi}(\x) = \sech{(\x)}^2 \sech{(n\,\tanh{(\x)})}^2.
\label{kinkssss}
\end{equation}

With the help of Eqs.~(\ref{hyp00}), the other BPS deformed functions are obtained as
\begin{eqnarray}
y_{\phi}(\x) &=& \sech{(\x)}^2 \tanh{(n\,\tanh{(\x)})}\sech{(n\,\tanh{(\x)})},\nonumber\\
z_{\psi}(\x) &=& \sech{(\x)}^2 \tanh{(n\,\tanh{(\x)})}.
\label{kinksmmm}
\end{eqnarray}
These solutions and all the others depicted from the last three columns of Fig.~\ref{Fig04A} correspond to some kind of non-topological lump-like defect.
The plots show the results for the primitive defects, $\chi(\x)$, $w_{\chi}(\x) = d\chi/d\x$, and $\rho(\chi(\x))$; and for the corresponding deformed defects, $\phi^{(n)}(\x)$, $y_{\phi}(\x) = d\phi^{(n)}/d\x$, and $\rho(\phi^{(n)}(\x))$; $\varphi^{(n)}(\x)$, $x_{\varphi}(\x) = d\varphi^{(n)}/d\x$, and $\rho(\varphi^{(n)}(\x))$; and $\psi^{(n)}(\x)$, $z_{\psi}(\x) = d\psi^{(n)}/d\x$, and $\rho(\psi^{(n)}(\x))$.
We have set $n = k \pi /6$ with $k$ in the range between $1$ and $2$, by steps of $0.2$ units, in order to depict the analytical dependence on the free parameter $n$.

In case of assuming that the primitive $\lambda \chi^4$ kink solution from Eq.(\ref{s1}) triggers the $4$-cyclically deformed chains discussed above, the deformed defect structures have their explicit analytical dependence on $\x$ obtained by substituting $\chi_{1}(\x)$ into the results from Eqs.~(\ref{hyp}).
Through this explicit dependence on $\x$, topological charges (in case of kink-like solutions) and topological masses can be straightforwardly obtained.

The topological charge for the novel kink-like deformed defect obtained from the hyperbolically deformed structure described by Eqs.~(\ref{kink}-\ref{kinks}) is therefore given by
\begin{equation}
Q_{(n)}^{\varphi} = 2\frac{\tanh{(n)}}{n},~~\mbox{with} ~~ n > 0,
\label{kinks}
\end{equation}
and the topological masses analytically obtained as function of the free parameter $n$ are given by
\begin{eqnarray}
M^{\psi}_{(n)} &=& \frac{4}{3} + \frac{\pi^2}{6 n^3} - \frac{2}{n} - \frac{4}{n^2} \ln{\left[1 + \exp{(-2 n)}\right]} +\frac{2}{n^3} Li_{2}\left[-\exp{(-2n)}\right],\nonumber\\
M^{\phi}_{(n)} &=&  - \frac{\pi^2}{18 n^3} + \frac{2}{3n} +
\frac{4}{3 n^2} \ln{\left[1 + \exp{(-2 n)}\right]}\nonumber\\
&&~~~~~~~~
-\frac{2}{3 n^3} \left\{ Li_{2}\left[-\exp{(-2n)}\right] + \tanh{(n)} - n \sech{(n)}^2\right\},\nonumber\\
M^{\varphi}_{(n)} &=&  - \frac{\pi^2}{9 n^3} + \frac{4}{3n} + \frac{8}{3 n^2} \ln{\left[1 + \exp{(-2 n)}\right]}\nonumber\\
&&~~~~~~~~
-\frac{2}{3 n^3} \left\{ 2 Li_{2}\left[-\exp{(-2n)}\right] - \tanh{(n)} + n \sech{(n)}^2\right\}.
\label{mass}
\end{eqnarray}

\subsection{Trigonometric Deformation}

Let us now consider the set of auxiliary functions described by
\begin{eqnarray}
\psi^{(n)}_{\chi} &=& -\sin{(n\,\chi)}, \nonumber\\
\phi^{(n)}_{\chi} &=&  -\frac{\sin{(2\,n\,\chi)}}{2},\nonumber\\
\varphi^{(n)}_{\chi} &=&  \cos{(n\,\chi)}^2,
\label{trig0}
\end{eqnarray}
which, for a given  $w_{\chi}$ substituted into Eq.(\ref{topo024B}), completes the $4$-cyclic chain, and upon straightforward integrations leads to
\begin{eqnarray}
\psi^{(n)}(\chi) &=& \frac{1}{n}\left[\cos{(n\,\chi)} - \cos{(n)}\right], \nonumber\\
\phi^{(n)}(\chi) &=& \frac{1}{4n}\left[\cos{(2n\,\chi)} - \cos{(2n)}\right], \nonumber\\
\varphi^{(n)}(\chi) &=&  \frac{1}{4n}\left[2 n\, \chi + \sin{(2n\,\chi)}\right],
\label{trig}
\end{eqnarray}
with constants chosen to follow the same criteria as for the previous solutions.
From Eq.(\ref{hypA3B}), one straightforwardly identifies the closure relation

\begin{eqnarray}
w_{\chi}^{2}
&=& w_{\chi}^{2}\left[\sin{(n\,\chi)}^2 + \cos{(n\,\chi)}^2\left(\sin{(n\,\chi)}^2 + \cos{(n\,\chi)}^2\right)\right]\nonumber\\
&=& w_{\chi}^{2}\left[\psi^{(n)\,2}_{\chi} + \phi^{(n)\,2}_{\chi} + \varphi^{(n)\,2}_{\chi}\right]\nonumber\\
&=& z_{\psi}^{2} + y_{\phi}^{2} + x_{\varphi}^{2},
\label{trig1}
\end{eqnarray}
which constrains the values of the topological masses obtained from trigonometrically deformed defects.

\begin{figure}
\centerline{\psfig{file= 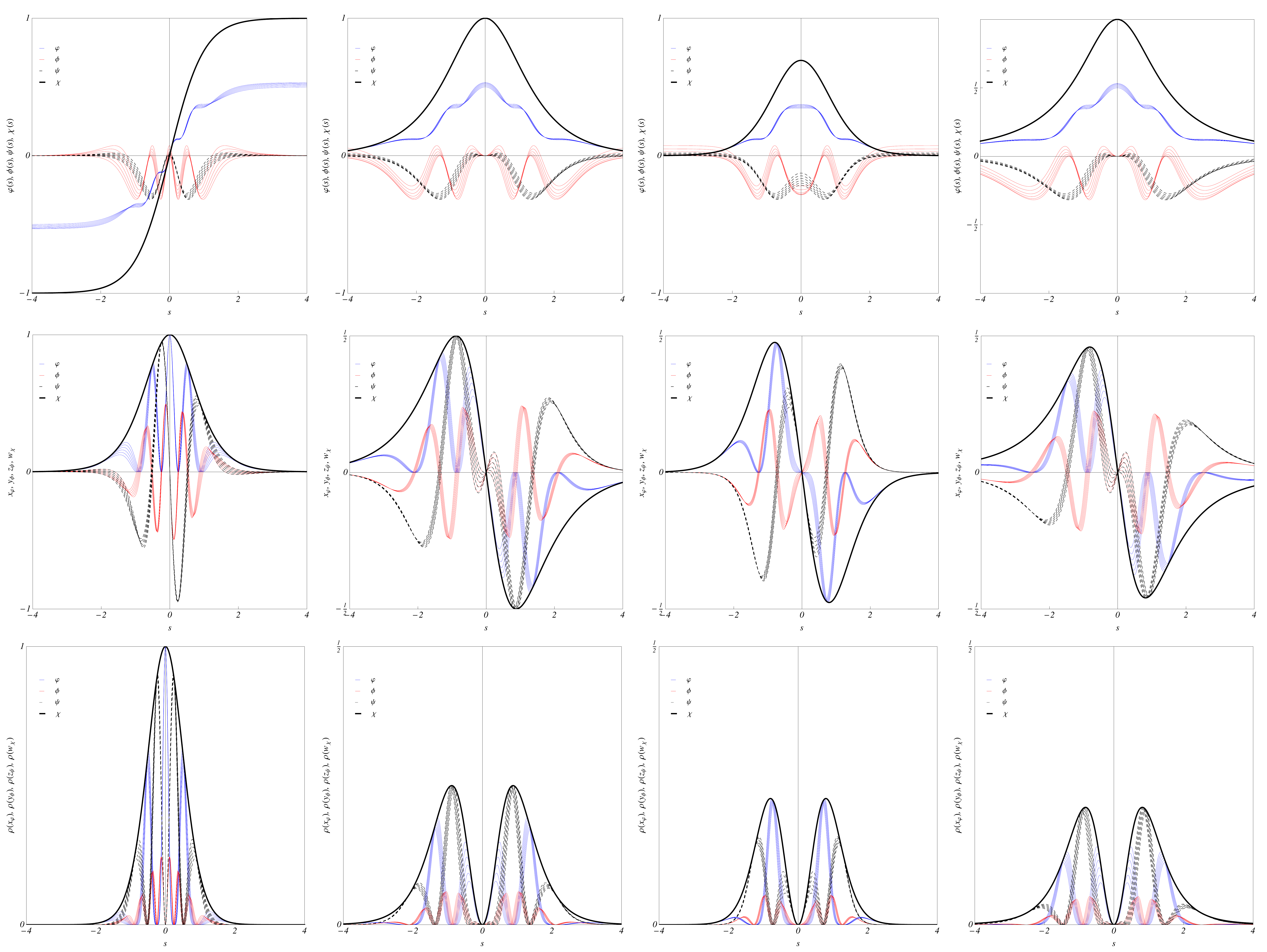, width= 18 cm}}
\caption{\small  The $4$-cyclically deformed defects obtained from {\em trigonometric}  deformation functions.
Results from each column are again triggered by solutions for $\chi(\x)$, in correspondence with Eqs.(\ref{s1}-\ref{s4}).
They are in correspondence with those from Fig.~(\ref{Fig04A}).
We have set $n = k \pi /6$ with $k$ in the range between $12$ and $13$, by steps of $0.2$ units, in order to depict the oscillatory analytical profile related to $n$, since for $n\lesssim1$ these plots perturbatively approximate those from Fig.~(\ref{Fig04A}).}
\label{Fig04B}
\end{figure}

Fig.~\ref{Fig04B} reproduces the analytical dependence on $\x$ for the $4$-cyclically deformed defects produced by the trigonometric deformation functions introduced above.
Four different cases are considered in correspondence with the primitive $\chi$ defect through Eqs.(\ref{s1}-\ref{s4}), where again the last three cases are merely illustrative.
All the cases depicted from Fig.~\ref{Fig04B} are in correspondence with those from Fig.~\ref{Fig04A} for hyperbolic deformations.
The unique kink-like defect, in this case, is given by
\begin{equation}
\varphi^{(n)}(\x) = \sin{(n\,\tanh{(\x)})}^{2},
\label{kinkt}
\end{equation}
with
\begin{equation}
x_{\varphi}(\x) = \sech{(\x)}^2 \sin{(n\,\tanh{(\x)})}^2.
\label{kinktrig}
\end{equation}
Eqs.~(\ref{trig0}) lead to the others BPS deformed lump-like solutions,
\begin{eqnarray}
y_{\phi}(\x) &=& \sech{(\x)}^2 \sin{(n\,\tanh{(\x)})}\cos{(n\,\tanh{(\x)})},\nonumber\\
z_{\psi}(\x) &=& \sech{(\x)}^2 \sin{(n\,\tanh{(\x)})},\label{kinkstrig}
\end{eqnarray}
once one has assumed $\chi$ from Eq.(\ref{s1}).
All the others solutions depicted from Fig.~\ref{Fig04B} correspond to some kind of non-topological lump-like defect.

In order to identify and emphasize the oscillatory behavior of such novel deformed defects, we have set $n = k \pi /6$ with $k$ in the range between $12$ and $13$, by steps of $0.2$ units, through which the analytical dependence on the free parameter $n$ can be depicted.
One can infer that, for small values of $n$, namely $n < 2$, the oscillatory pattern of trigonometric deformation disappears, and hyperbolic and trigonometric deformations are close to each other.

To sum up, once the primitive $\lambda \chi^4$ kink solution from Eq.(\ref{s1}) triggers the $4$-cyclically deformed chain discussed above, topological charges (in case of kink-like solutions) topological masses can be straightforwardly obtained
through this explicit dependence on $\x$.
The topological charge for the novel kink-like deformed defects obtained from the trigonometrically deformed structure described by Eqs.~(\ref{kinktrig}-\ref{kinkstrig}) is given by
\begin{equation}
Q_{(n)}^{\varphi} = \frac{n +\sin{(n)}\cos{(n)}}{n} ,
\label{kinksstrig}
\end{equation}
and the topological masses are given by
\begin{eqnarray}
M^{\psi}_{(n)} &=& \frac{2}{3} + \frac{2n \cos{(2n)} - \sin{(2n)}}{4 n^3},\nonumber\\
M^{\phi}_{(n)} &=& \frac{1}{6} + \frac{4n \cos{(4n)} - \sin{(4n)}}{128 n^3},\nonumber\\
M^{\varphi}_{(n)} &=& \frac{1}{2} - \frac{64n \cos{(2n)} +  4n \cos{(4n)} - 32 \sin{(2n)} - \sin{(4n)}}{128 n^3}.
\label{masstrig}
\end{eqnarray}

The topological charge for the novel kink-like deformed defects related to $\varphi^{(n)}$ and obtained from hyperbolic and trigonometric deformations in dependence on the free parameter $n$ can be depicted from Fig.~\ref{Fig05C}.
It also includes the results for the topological charge of the kink-like solution related to $\phi^{(n)}$ obtained from the $3$-cyclically deformed chain discussed in the previous section.

\begin{figure}
\centerline{\psfig{file= 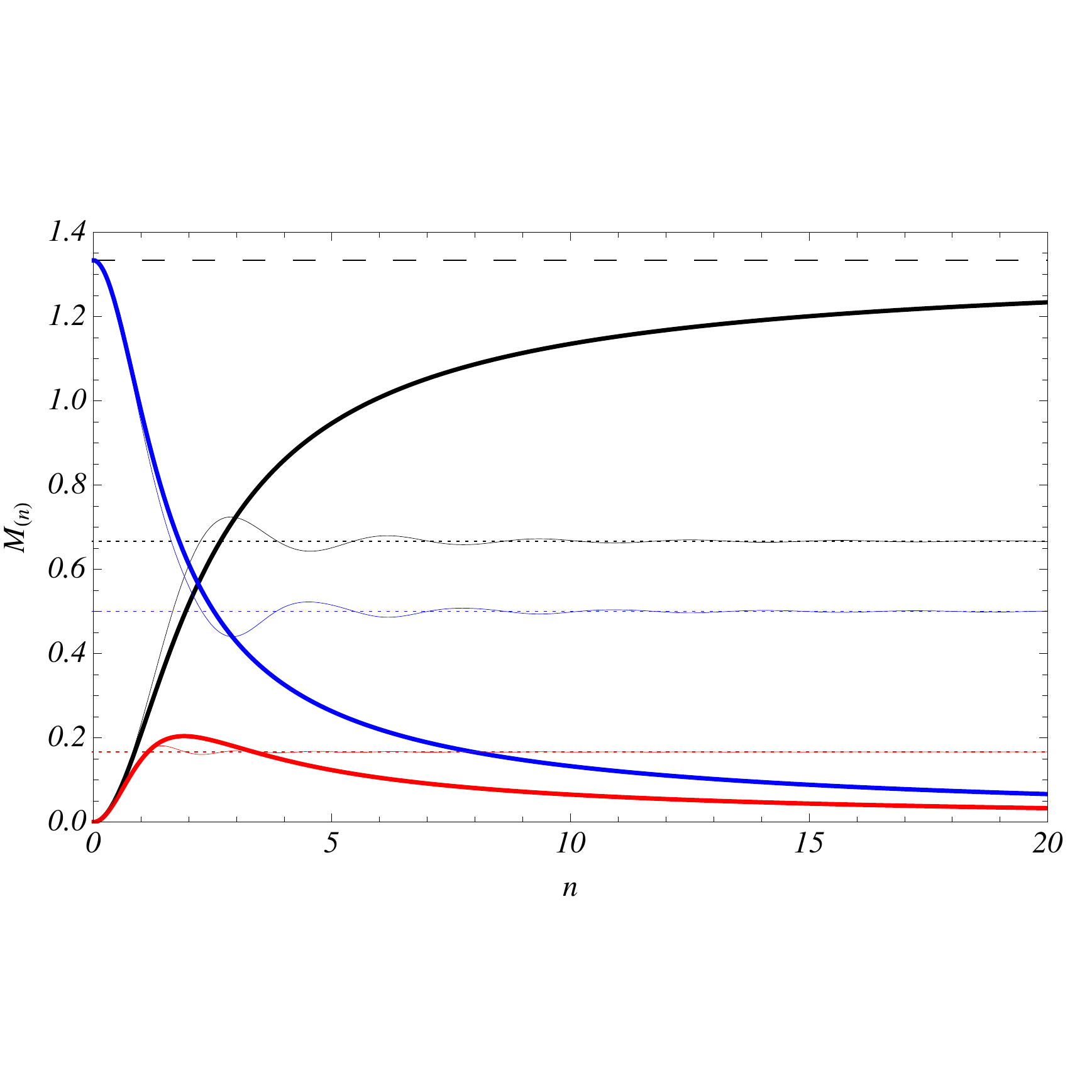, width= 14 cm}}
\caption{\small  Total energy of localized solutions (topological masses, $M_{(n)}$) as function of the parameter $n$ for hyperbolic (thick lines) and trigonometric (thin lines) $4$-cyclic deformation chains.
Results are for $M^{\varphi}_{(n)}$ (blue lines), $M^{\phi}_{(n)}$ (red lines), and $M^{\psi}_{(n)}$ (black lines).}
\label{Fig04C}
\end{figure}

\begin{figure}
\centerline{\psfig{file= 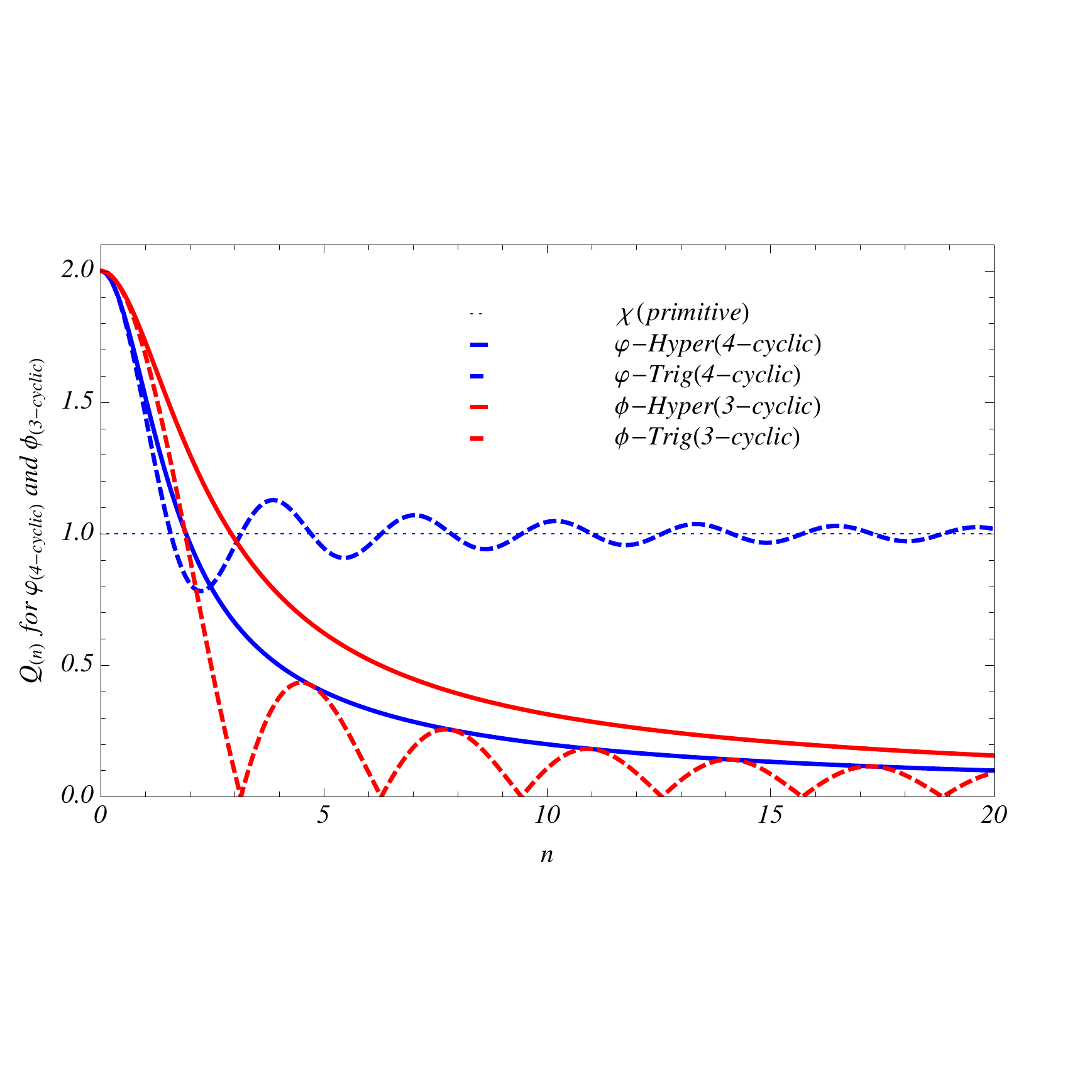, width= 14 cm}}
\caption{\small  Topological charges, $Q_{(n)}$, as function of the parameter $n$ for hyperbolic (thick lines) and trigonometric (dashed lines) deformation chains.
Results are for the kink-like solutions obtained from $4$-cyclic (blue lines) and $3$-cyclic (red lines) deformation chains respectively related to $\varphi^{(n)}$ and $\phi^{(n)}$.}
\label{Fig05C}
\end{figure}

Finally, the topological masses analytically obtained as functions of the free parameter $n$, for deformed defect structures cyclically related to the primitive $\lambda \chi^4$ kink solution  (c. f. Eq.~(\ref{s1}) and the first columns of Figs.~\ref{Fig04A} and \ref{Fig04B}) are depicted from Fig.~\ref{Fig04C}.
It indeed ratifies the most relevant result of our technique.
It is related to the constraint relations determined through Eqs.~(\ref{hyp1}) and (\ref{trig1}) respectively for hyperbolic and trigonometric deformation functions.
From Fig.~\ref{Fig04C}, one can verify forthwith that
\begin{equation}
M^{(n)}_{\psi} + M^{(n)}_{\varphi} + M^{(n)}_{\phi}  = M_{\chi} = \frac{4}{3},
\end{equation}
in both cases which, in the asymptotic limit, $n\rightarrow \infty$, leads to
\begin{equation}
\frac{4}{3} + 0 + 0 =  \frac{4}{3},
\end{equation}
for hyperbolic deformations, and to
\begin{equation}
\frac{2}{3} + \frac{1}{6} +\frac{1}{2} =  \frac{4}{3},
\end{equation}
for trigonometric deformations.

\section{N-cyclic deformations}

Departing from a triggering defect given by $\chi$ (eventually related to the $\lambda \chi^4$ kink solution), the observation of the chain rule constrained by hyperbolic and trigonometric fundamental relations allows one to investigate the possibility of obtaining $N$-cyclic deformations.

Let us finally introduce a generalized $s$-deformation operation parameterized by the $s$-derivative given by
\begin{equation}
g^{[s]}(\phi^{[s]}) = \frac{d\phi^{[s]}}{d\phi^{[s-1]}},
\end{equation}
where $\phi^{[s]}$, with $s = 1,\,2,\, \ldots,\,N$, are real scalar fields describing defect structures generated from a $N$-cyclically deformation chain triggered by a primitive defect $\chi\equiv\chi(\x)$, such that one can define $\phi^{[s]} \equiv \phi^{[s]}(\chi)$.

The hyperbolic deformation chain shall be given in terms of
\begin{eqnarray}
\phi^{[0]}_{\chi} &=& \tanh{(\chi)}, \nonumber\\
\phi^{[1]}_{\chi} &=& \tanh{(\chi)}\sech{(\chi)}, \nonumber\\
\phi^{[2]}_{\chi} &=& \tanh{(\chi)}\sech{(\chi)}^{2}, \nonumber\\
\vdots \;\;\,&=&\qquad\qquad\vdots \nonumber\\
\phi^{[N-1]}_{\chi} &=& \tanh{(\chi)}\sech{(\chi)}^{N-1}, \nonumber\\
\phi^{[N]}_{\chi} &=& \sech{(\chi)}^{N},
\label{AA}
\end{eqnarray}
which upon straightforward integrations results into
\begin{eqnarray}
\phi^{[0]}(\chi) &=& \ln{[\cosh{(\chi)}]}, \nonumber\\
\phi^{[1]}(\chi) &=& - \sech{(\chi)}, \nonumber\\
\phi^{[2]}(\chi) &=& - \frac{1}{2} \sech{(\chi)}^{2}, \nonumber\\
\vdots \;\;\,&=&\qquad\vdots \nonumber\\
\phi^{[N-1]}(\chi) &=& - \frac{1}{N-1}\sech{(\chi)}^{N-1}, \nonumber\\
\phi^{[N]}(\chi) &=&  \sinh{(\chi)}\, _2F_1\left[\frac{1}{2},\, \frac{1+N}{2},\, \frac{3}{2},\, -\sinh{(\chi)}^2\right], \nonumber\\
\label{BB}
\end{eqnarray}
where $_2F_1$ is the Gauss' hypergeometric function, and we have not considered any adjust due to integration constants.
From Eqs.(\ref{AA}) one easily identifies that
\begin{equation}
g^{[s]}(\phi^{[s]}) = \sech{(\chi)} \equiv \exp{[-\phi^{[0]}]},~~ s = 1, 2,\,\ldots,\,N-1,
\end{equation}
and
\begin{equation}
g^{[N]}(\phi^{[N]}) = \sinh{(\chi)}^{-1},
\end{equation}
such that
\begin{equation}
\prod_{s=1}^{N-1}{g^{[s]}(\phi^{[s]})} = \frac{d\phi^{[N-1]}}{d\phi^{[0]}} = \sech{(\chi)}^{(N-1)} \equiv \exp{[- (N-1) \phi^{[0]}]},
\label{DD}
\end{equation}
from which a very simplified form for the chain rule of the $N$-cyclic deformation can be obtained as
\begin{equation}
\frac{d\phi^{[0]}}{d\chi}\,\prod_{s=1}^{N}{g^{[s]}(\phi^{[s]})}\,\frac{d\chi}{d\phi^{[N]}}=
\frac{d\phi^{[0]}}{d\chi}\,\left(\prod_{s=1}^{N-1}{g^{[s]}(\phi^{[s]})}\right)\,g^{[N]}(\phi^{[N]})\,\frac{d\chi}{d\phi^{[N]}}= 1.
\label{EE}
\end{equation}

Upon introducing the generalized BPS functions,
\begin{eqnarray}
y^{[s]}_{\phi^{[s]}} &=&  y^{[s-1]}_{\phi^{[s-1]}}\, g^{[s]}(\phi^{[s]}) = \frac{d \phi^{[s]}}{d\x}
= \frac{d \chi}{d\x} \frac{d \phi^{[s]}}{d\chi} =  w_{\chi} \phi^{[s]}_{\chi},
\end{eqnarray}
with $w_{\chi} = d\chi/d\x$, it is straightforwardly verified that
\begin{eqnarray}
\sum_{s=0}^{N-1}{(y^{[s]}_{\phi^{[s]}})^2}
&=&  w_{\chi}^2 \sum_{s=0}^{N-1}{(\phi^{[s]}_{\chi})^2}
= w_{\chi}^2 \tanh{(\chi)}^2\,\sum_{s=0}^{N-1}{(\sech{(\chi)})^{2s}}\nonumber\\
&=&  w_{\chi}^2 \tanh{(\chi)}^2\,\frac{1 - \sech{(\chi)}^{2N}}{1-\sech{(\chi)}^2} = w_{\chi}^2 \left( 1 - \sech{(\chi)}^{2N}\right)\nonumber\\
&=& w_{\chi}^2 \left[ 1 - (\phi^{[N]}_{\chi})^2 \right] = w_{\chi}^2 - (y^{[N]}_{\phi^{[N]}})^2,
\label{BBBB}
\end{eqnarray}
which results into the constraint equation,
\begin{eqnarray}
\sum_{s=0}^{N}{(y^{[s]}_{\phi^{[s]}})^2} &=& w_{\chi}^2.
\label{cd1}
\end{eqnarray}

Consequently the topological masses are constrained by
\begin{eqnarray}
\sum_{s=0}^{N}{M^{\phi^{[s]}}} &=& M^{\chi}.
\label{cd2}
\end{eqnarray}

The trigonometric deformation chain can be exactly obtained by substituting $\tanh{(\chi)}$ by $-\sin{(\chi)}$ and
$\sech{(\chi)}$ by $\cos{(\chi)}$ into Eqs.(\ref{BB}).
It shall result into
\begin{eqnarray}
\phi^{[0]}(\chi) &=& \cos{(\chi)}, \nonumber\\
\phi^{[1]}(\chi) &=& \frac{1}{2}\cos{(\chi)}^{2}, \nonumber\\
\phi^{[2]}(\chi) &=& \frac{1}{3}\cos{(\chi)}^{3}, \nonumber\\
\vdots \;\;\,&=&\qquad\vdots \nonumber\\
\phi^{[N-1]}(\chi) &=& \frac{1}{N}\cos{(\chi)}^{N}, \nonumber\\
\phi^{[N]}(\chi) &=&  -\frac{\cos{(\chi)^{N+1}}}{N+1}\, _2F_1\left[\frac{1+N}{2},\, \frac{1}{2},\, \frac{3 + N}{2},\, \cos{(\chi)}^2\right],
\label{BB2}
\end{eqnarray}
and into the same set of constraints described by Eqs.~(\ref{cd1}-\ref{cd2}).

Essentially, the $N$-finite set of bijective hyperbolic/trigonometric functions with non-vanishing asymptotically bounded derivatives provides us with the necessary and sufficient tools for defining the $N$-cyclic deformation chain that systematically results into constraints involving the topological mass of deformed structures.

\section{Conclusions}

We have introduced and scrutinized a systematic procedure for obtaining defect structures through cyclic deformation chains involving topological and non-topological solutions in models described by a single real scalar field.
After embedding some previously investigated deformed defect structures, namely the {\em sine-Gordon} kink, and the {\em bell}-shaped lump-like structure, into a $3$-cyclic deformation chain supported by the primitive kink solution of the $\lambda \phi^4$ theory, we have investigated the existence of a systematic technique to obtain novel defect structures.
As a preliminary study, we have discussed the kink-like solution engendered by a {\em plateau}-shaped potential that is generated from a logarithmic dependent deformation of the $\lambda \phi^4$ theory through an arbitrary (non-systematic) cyclic deformation chain.
It subtly mimics the slow-rolling scalar field potentials at $1+1$ dimensions.

In a subsequent step, we have verified that a set of rules for constructing constraint relations involving topological masses of cyclically generated defects could be systematically established.
The main goal of our study was concerned with the possibility of recovering some primitive defect structures through a regenerative, unidirectional, and eventually irreversible, sequence of deformation operations via hyperbolic and trigonometric functions.
We have verified that the cyclic deformation device can simultaneously support kink-like and lump-like defects into $3$- and $4$-cyclic deformation chains with topological mass values analytically obtained, and constraint equations established.
The possibility of a straightforward generalization of the analytical calculations to $N$-cyclic deformations was demonstrated. The perspective of applying the regeneration mechanism to more complex primitive triggering defects, for instance, the {\em sine-Gordon} kink-like solution, connecting it to novel defect structures, may be considered in subsequent analysis.

We  have further emphasized that several components of the cyclic chains that we have studied correspond to lump-like solutions.
Since they are non-monotonic functions, they induce some subtle features to the non-topological structures.
As pointed out by some preliminary issues \cite{Bas01,Bas02,Bas03}, for non-topological structures, namely the lump-like ones, the quantum mechanical problem engenders zero-mode functions with one or more nodes.
It leads to the possibility of negative energy bound states, providing us  unstable solutions, and creating additional difficulties for determining such bound states through the related quantum mechanics.

Whatever the outcome of these investigations, there is no doubt that the study of defect structures through a systematic process of producing $N$-cyclic deformations leads, at least, to the speculation of novel scenarios of physical and mathematical applicability of defect structures, for instance, at some kind of cyclically connected optical system, or some cyclically regenerative sequence of phase transitions.

{\em Acknowledgment: This work was supported by the Brazilian Agency CNPq (grant 300233/2010-8 and grants 476580/2010-2 and 304862/2009-6.}

\renewcommand{\baselinestretch}{1.0}

\end{document}